\documentstyle[aps,manuscript,psfig,prb]{revtex}
\input epsf
\begin{document}

\title{Equation of state and critical behavior of polymer models:
A quantitative comparison between Wertheim's thermodynamic perturbation theory and computer simulations.
}

\author{L. Gonz\'alez MacDowell $^{\dagger}$, M. M{\"u}ller $^{\ddagger}$, 
	  C. Vega $^{\dagger}$, and K. Binder $^{\ddagger}$   \\
	  $\dagger$ Dpto. de Qu\'{\i}mica F\'{\i}sica, 
			Facultad de C.C. Qu\'{\i}micas, \\ 
			Universidad Complutense, Madrid 28040, Spain \\
        $\ddagger$ Institut f{\"u}r Physik, WA 331, 
			Johannes Gutenberg Universit{\"a}t \\
			D-55099 Mainz, Germany}

\maketitle

\begin{abstract}
We present an application of Wertheim's Thermodynamic Perturbation Theory
(TPT1) to a simple coarse grained
model made of flexibly bonded Lennard-Jones monomers.
We use both the Reference Hyper-Netted-Chain (RHNC) and Mean Spherical 
approximation (MSA) integral equation theories to describe the properties
of the reference fluid. The equation of state, the density dependence of the
excess chemical potential, and
the critical points of the liquid--vapor transition are compared with simulation results and good
agreement is found. The RHNC version is somewhat more accurate, while
the MSA version has the advantage of being almost analytic. We analyze
the scaling behavior of the critical point of chain fluids according
to TPT1 and find it to reproduce the mean field exponents:
The critical monomer density is predicted
to vanish as  $n^{-1/2}$  upon increasing the chain length $n$ while the 
critical temperature is predicted
to reach an asymptotic finite temperature that is attained as $n^{-1/2}$.
The predicted asymptotic finite critical temperature obtained from
the RHNC and MSA versions of TPT1 is found to be in good agreement
with the $\Theta$ point of our polymer model as obtained from the
temperature dependence of the single chain conformations.
\end{abstract}

\section{introduction}

For a long time, the prediction of the equation of state of
polymers from first principles was such a difficult task that
it could not be solved without the need of very idealized
models with more or less meaningful empirical parameters.
Perhaps the most successful of the early attempts was that
of Prigogine {\em et al.} \cite{prigogine57} who considered a
lattice and introduced an ad-hoc empirical parameter known
as the ``number of degrees of freedom per monomer''. Later on,
Flory {\em et al.} \cite{flory64a,flory64b,flory64c} extended 
this theory to the continuum at the
cost of introducing some additional parameters, leading  to
what is now known as the FOVE theory. 
Another approach to the problem is based on the polymer+solvent and
polymer+vacuum analogy. In this way, the well known Flory-Huggins
\cite{flory41,huggins41} (FH) and Sanchez-Lacombe \cite{sanchez76} equations of
state have also been employed to describe the behavior of pure fluids.

On the other hand, the approach from liquid state theory
has taken much longer to yield useful results but has now
reached a point where a rather satisfactory description of the
equation of state of idealized polymer models in the continuum
is affordable without the need of any empirical parameters whatsoever.
The most popular approaches are the 
Polymer Reference Interaction Site Model (PRISM) of Curro and
Schweizer, \cite{curro87} the Generalized Dimer Flory theory (GDF)
of Honnell and Hall \cite{honnell89} and the 
Thermodynamic Perturbation Theory of Wertheim (TPT1).
\cite{wertheim87} 
The latter has been widely used not only because it yields results
that are of similar or superior quality than other alternatives,
but because it is the simplest and more tractable of all them and
demands a minimum of information.

Originally, this theory was developed to consider fluids of associating
hard spheres. \cite{wertheim84a,wertheim84b,wertheim86a,wertheim86b} 
Later on, it was
realized that in the limit of infinite associating strength, a 
polydisperse mixture of polymers was recovered. \cite{wertheim87}
Chapman {\em et al.} extended the theory to monodisperse polymer fluids
and rewrote the equations in a very convenient notation. \cite{chapman88}
Later, by adding a mean field perturbative contribution, Jackson {\em et al.}
showed that the theory was able to describe the behavior of very
different sorts of real fluids. \cite{jackson95} 
Meanwhile, it was realized that the theory could be used just as well
to consider polymers made of attractive monomers.\cite{chapman90,chapman90b} 
In this way, it
is possible to describe real fluids with improved accuracy and no need  
for a mean field--like perturbation. 
Since then, the theory has achieved enormous popularity and has been
applied to describe polymers of tangent Lennard-Jones beads
\cite{banaszak94,johnson94,escobedo96,blas97}
and square wells, \cite{banaszak93,gil-villegas97} as well as to describe 
real fluids in chemical
engineering applications. \cite{huang90,chen98,blas98} Furthermore, 
modifications
of the original theory have been proposed that allow to describe
realistic polymer models without the need of empirical
parameters. \cite{vega94,vega96b}

A very interesting issue both from the practical and theoretical
point of view is the behavior of the critical point of polymer
fluids as the number of monomers increases. By invoking the
polymer+solvent and polymer+vacuum analogy, one would expect from
the FH theory that the pure polymer fluid should reach an asymptotic
critical temperature, whereas the critical mass density should become 
vanishingly
small. However, in a recent paper, Chatterjee and Schweizer 
\cite{chatterjee98} have pointed out that this analogy cannot be
taken for granted because the FH scaling predictions 
are determined by imposing equal chemical potentials, whereas
the critical point of pure polymer fluids is related to a phase
equilibria that results from the condition of equal pressure at
a given temperature. \cite{rowlinson82}

On the other hand, based on rather limited amount of data,
several empirical correlations have been employed to predict
the critical properties of substances such as polymethylene.
\cite{tsonopoulos87,tsonopoulos93,nakanishi60,korsten98}
These correlations have predicted  widely different behavior,
ranging from infinite critical temperature \cite{korsten98} 
to finite asymptotic
critical mass density. \cite{tsonopoulos87,nakanishi60,korsten98}
More soundly based equations such as the FOVE have been recently
employed to support the idea that the critical mass density
could reach an asymptotic constant value. \cite{tsonopoulos93} 
However, such an
approach relies on extending the applicability of the FOVE theory
to densities below which it was not meant to be used. \cite{flory64a}

Renormalization group calculations, however, show that the FH
mean field theory yields the correct asymptotic dependence in the long
chain length limit\cite{dup82,dup87}, albeit this asymptotic behavior
is only reached for extremely long chains and the corrections to scaling 
are enormous even for typical chain length considered experimentally.\cite{hs99}

More recently, new experimental techniques have allowed to 
measure the critical points of longer n-alkanes whose critical
temperature lays above the point where thermal decomposition
starts. \cite{anselme90,nikitin94,nikitin97} These experiments show that 
the critical mass density reaches
a maximum and then starts to decrease. \cite{anselme90} Simulation results of
both realistic alkanes \cite{smit95a} and idealized polymer models
\cite{sheng94,wilding96} give support to this finding.

Surprisingly, very little attention has been devoted to the study
of this problem from the point of view of the modern theories
such as PRISM, GFD or TPT1. In a recent paper, the PRISM was
employed in an attempt to solve the question in the framework of an analytical tractable theory,
leading to different predictions depending on the closure employed to solve the
PRISM equation. \cite{chatterjee98}
Other previous studies using TPT1 plus a mean field attractive
contribution suggested that the critical mass density should  
vanish in the infinite chain length limit. \cite{vega96a} However, such a
conclusion relied on the assumption that the mean field 
contribution increases linearly with the chain length, a point
which at present cannot be taken for granted.
Nikitin {\em et al.} \cite{nikitin96,nikitin97} have recently presented 
a theory 
that may be considered as the simplest possible approximation to
Wertheim's theory and arrive  to the same conclusion as 
reference \cite{vega96a}.

In this paper we will try to reach further understanding of
this issue. In the next section, we will review the
fundamentals of Wertheim's theory by using  somewhat different but
more physically appealing arguments suggested by Zhou and Stell.  
\cite{zhou92} We will then show by means of  general arguments
how the scaling laws for the critical properties are closely
related to the virial coefficients (section III). 
We will then apply TPT1 to a polymer model and 
briefly describe how to implement the theory in section IV
and in the Appendixes.
Section V describes the details of the simulations performed
to test the theory. We then devote section VI to present our 
results and close with a brief conclusion.

\section{Equation of state of polymers}

\subsection{Preliminary definitions}

We consider a monodisperse fluid of polymers made
up of $n$ monomers each. Non bonded monomers of the
same polymer and monomers belonging to different 
molecules interact through some pair potential,
$u_0$, while adjacent monomers of the same molecule
have an additional bonding potential, $\Phi$, responsible
for the connectivity of the polymer. This potential is
such that its action vanishes beyond some well defined
inter-atomic distance. If eventually, two adjacent monomers
were found outside this range, they would no longer be
bonded and the n-mer would be considered to have broken.
In practice, this can be avoided by making the well
depth of the associating potential infinitely large.

Alternatively, we may  consider 
an associating multicomponent mixture of $n$ different
monomeric species, $A$, $B$, $C$, etc.
Each of these species interacts with members
of its own species and with members of the remaining
species by means of $u_0$. Furthermore, 
the bonding potential $\Phi$ is responsible for associating
reactions between monomers of type $A$ with monomers of
type $B$, monomers of type $B$ with monomers of type $C$
and so on. More concisely, the association reaction
taking place is of the form:
\begin{equation}
A+B+C+\cdots \rightleftharpoons ABC\cdots
\end{equation}
As the bonding potential is pairwise, the only requirement
that is needed to define the '$ABC\cdots$' complex as 
an n-mer is that each of the adjacent pairs be found
within the range of the bonding potential.
Clearly, for an equimolar composition of such a mixture
in the limit of complete association, we are lead to
a system identical to that described in the preceding
paragraph. From a physical point of view, this limit
is reached for infinite well depths. However,
in what follows it will prove useful to consider that
the well depth
has some arbitrary finite value. The system is then
made of a mixture of free monomers and n-mers whose
composition depends on the nature of the bonding potential.

Considering the similarity between the two systems described,
we may get an approximation for the equation of state for the
chain molecule by studying the  behavior of
the associating system. In order to do so, we will
obtain an expression for the free energy in terms of the
degree of association of the mixture. We will then 
relate the degree of association to the structure of
the system and introduce some simplifying assumptions
for the n-body correlations. Finally, we will take the
limit of complete association and get an equation of
state for the chain fluid. 

\subsection{The association reaction}

Let us consider an associating system as that described
before, which is initially prepared by mixing in equal
proportions the pure monomers, so that there are $N$
monomers of each species inside a volume $V$ and the
resulting number density
of each of the species is $\rho$. The system will eventually
reach a state of equilibrium, whereby a fraction of the
monomers of each species has associated to form n-mers. Let this 
fraction be $\alpha$. Then, the remaining concentration
of non bonded monomers of each species is given by $\rho(1-\alpha)$, while,
according to the stoichiometry of the reaction,
the concentration of n-mers will be $\rho \alpha$ (note that in the
limit of complete association $\rho$ will actually designate the polymer
number density).
Schematically, the process can be described as follows:
\begin{equation}
\begin{tabular}{cccccccccc}
& &$A$ & $+$ & $B$ & + & $C$ & $\cdots$ & $\rightleftharpoons$ & $ABC\cdots$ \\ 
initially & & $\rho$ & & $\rho$ & & $\rho$  & & & $0$ \\
at equilibrium & & $\rho(1-\alpha)$ & & $\rho(1-\alpha)$ & & $\rho(1-\alpha)$ & & & $\rho\alpha$ \\
\end{tabular}
\nonumber
\end{equation}

Obviously, the number of n-mers formed is $V \rho \alpha = N \alpha$,
while the number of remaining free monomers of each species is
$N(1-\alpha)$. The total Helmholtz free energy of the system,
$A=G-pV$ is
therefore given by:
\begin{equation}
\label{eqA1}
A(\alpha) = \sum_{i=1}^n N(1-\alpha)\mu_i(\alpha) + N\alpha \mu_{\rm n-mer}
									   (\alpha)
 -p(\alpha)V
\end{equation}  

In what follows, we associate a number from $1$ to $n$ to each of
the n species. Therefore, $\mu_i$ stands for the chemical potential
of species $i$ in the mixture of composition given by $\alpha$. 

It is important to realize that the composition of the mixture
as given by the association degree is appropriate to a given model
potential. In the above expression for the free energy, we have considered
a system of monomers that interact through some reference potential,
$u_0$, while a bonding potential is responsible for the connectivity of
the n-mer. Let us now consider a reference fluid
made of monomers that interact through the reference potential but
that have no association potential whatsoever. In this case, the 
free energy is best expressed in terms of the chemical potential 
of the monomers alone. For the sake of simplicity, we will denote
this free energy as $A(\alpha=0)$, albeit from a strictly geometrical
point of view, n-mers could form eventually even in the absence of
a bonding potential:
\begin{equation}
\label{eqA2}
A(\alpha=0) = \sum_{i=1}^n N\mu_i(\alpha=0) - p(\alpha=0)V
\end{equation}

The free energy of the associating system measured relative to
that of the reference fluid is then given by:
\begin{equation}
\label{eqDeltaA_1}
\frac{\Delta A}{N} = \sum_{i=1}^n[\mu_i(\alpha) - \mu_i(\alpha=0)]
                   + \alpha[\mu_{\rm n-mer}(\alpha) - \sum_{i=1}^n 
									\mu_i(\alpha) ]
		   - [p(\alpha) - p(\alpha=0)]\frac{V}{N}
\end{equation}
This equation can be further simplified
by invoking the condition of chemical equilibrium of the
associating mixture, which reads:
\begin{equation}
\label{eqchemeq}
\sum_{i=1}^n \mu_i(\alpha)  - \mu_{\rm n-mer}(\alpha) = 0
\end{equation}
Substitution of the above equation into Eq. \ref{eqDeltaA_1} yields: 
\begin{equation}
\label{eqDeltaA_2}
\frac{\Delta A}{N} = \sum_{i=1}^n[\mu_i(\alpha) - \mu_i(\alpha=0)]
                   - [p(\alpha) - p(\alpha=0)]\frac{V}{N}
\end{equation}

Let us now express the chemical potentials of each of the components
in terms of an ideal and an excess contribution:
\begin{equation}
\label{eqmudiv}
\mu_i(\alpha) = \mu^{\rm id}_i(\alpha) + \mu^{\rm ex}_i(\alpha)
              = \mu_i^0 + kT\ln\rho_i(\alpha) + \mu_i^{\rm ex}(\alpha)
\end{equation}
Introducing the above expression for the chemical potential into
Eq. \ref{eqDeltaA_2}, we get:
\begin{equation}
\label{eqDeltaA_3}
\frac{\Delta A}{N} =  kT \ln \prod_{i=1}^n \frac{\rho_i(\alpha)}{\rho_i(\alpha=0)}
	           +  \sum_{i=1}^n[\mu_i^{\rm ex}(\alpha) - \mu_i^{\rm ex}(\alpha=0)]
                   - [p(\alpha) - p(\alpha=0)]\frac{V}{N}
\end{equation}

If we now recall that $\rho_i(\alpha)=\rho(1-\alpha)$ and 
$\rho_i(\alpha=0)=\rho$, we finally
obtain:
\begin{equation}
\label{eqDeltaA_4}
\frac{\Delta A}{N} =  n kT \ln (1-\alpha)
                   +  \sum_{i=1}^n[\mu_i^{\rm ex}(\alpha) - \mu_i^{\rm ex}(\alpha=0)]
                   - [p(\alpha) - p(\alpha=0)]\frac{V}{N}
\end{equation}
This is an exact equation for the difference in free energy between the
associating system (with $\alpha N$ n-mers) and the reference non-associating 
fluid of free monomers.
As it stands, it may seem rather useless, because it is a function
of the unknown quantities: $\alpha$, $\mu_i^{\rm ex}(\alpha)$
and $p(\alpha)$. However, we shall see that in the low density limit,
the above equation becomes a function of the degree of association
only and that this, in turn, may be obtained from knowledge of 
n-body correlations in the fluid. In a further approximation, the
n-body correlations of the fluid will be expressed in terms of
n-body correlations of a reference fluid of non--bonded monomers.
Finally, by invoking a superposition approximation, we will express
the n-body correlations in terms of two body correlations and
obtain an expression for the pressure that depends solely
on known quantities of a reference fluid of spherical particles.   

\subsection{The association reaction in the limit of low density}

In the limit of low density, the equation of state of the associating
system will depend only on the total number of particles in the fluid,
$N(\alpha)$:
\begin{equation}
\frac{pV}{kT} = N(\alpha)
\end{equation}
$N(\alpha)$ is simply obtained by summing the number of particles
of each species:
\begin{equation}
N(\alpha) = \sum_{i=1}^n N(1-\alpha)  + N\alpha = nN(1-\alpha) + N\alpha
\end{equation}

On the other hand, the number of particles of the completely un-associated
system is simply $N(\alpha=0)=nN$.
Therefore, the difference in pressure between the non--bonded system
and the system with degree of association $\alpha$ is:
\begin{equation}
\Delta p(\alpha) \frac{V}{N} = -kT \alpha (n-1)
\end{equation}
By using this expression and considering that, by definition, the
excess chemical potentials vanish in the limit of low density,
Eq. \ref{eqDeltaA_4} becomes:
\begin{equation}
\frac{\Delta A}{kT N} = n \ln (1-\alpha) + \alpha (n-1)
\end{equation}
This is an exact equation for the difference in free energy
in the limit of low density. Applying
the standard thermodynamic relationship connecting the pressure
with the free energy yields the exact expression for the difference 
in pressure in the limit of low density:
\begin{equation}
\label{eqP_1}
\frac{\Delta p}{kT \rho} = \rho \frac{\alpha(1-n)-1}{1-\alpha}
				\frac{\partial \alpha}{\partial \rho}
\end{equation}

In what follows, we will consider this expression to be valid in
all the density range. The search of an equation of state for
the associating system will be thus accomplished if we find an
expression for the association degree in terms of known properties.
Once this relation has been found, we will obtain an equation of
state for the system of n-mers by taking the limit of complete
association, i.e., $\alpha=1$.

\subsection{Relation between the degree of association and the structure
	    of the fluid}

\subsubsection{Expression for the degree of association in terms of
	       the excess chemical potential of the components}

First, consider the equilibrium constant of the reaction, defined
as the ratio of the concentration of the products to that of
the reactants:
\begin{equation}
\label{eqKeq}
K = \frac{ \alpha\rho }{ (1-\alpha)^n\rho^n } 
\end{equation}

The connection of the equilibrium constant to the thermodynamics
of the process may be obtained by expressing the chemical
potential of each of the components as in Eq. \ref{eqmudiv} and
substituting into the condition of chemical equilibrium,
Eq. \ref{eqchemeq}.
After some simple algebraic manipulations, we are lead to 
the following expression for the equilibrium constant:
\begin{equation}
\label{eqK_1}
kT\ln K + \mu^0_{\rm n-mer} - \sum_{i=1}^n \mu_i^0 + 
					\mu_{\rm n-mer}^{\rm ex}(\alpha)
       - \sum_{i=1}^n \mu_i^{\rm ex}(\alpha) = 0
\end{equation}
This expression may be further simplified by considering that, in
the limit of low densities, the excess chemical potentials vanish.
As a consequence of this, the low density equilibrium constant,
$K_0$, is given as follows:    
\begin{equation}
kT\ln K_0 = \sum_{i=1}^n \mu_i^0 - \mu^0_{\rm n-mer} 
\end{equation}
Substitution of this expression into Eq. \ref{eqK_1}, leads
finally to a simple equation for the equilibrium constant in
terms of the excess chemical potential of the components of the mixture:
\begin{equation}
\label{eqalfatermo}
kT \ln \frac{K}{K_0} = \sum_{i=1}^n \mu_i^{\rm ex}(\alpha) - 
				  \mu_{\rm n-mer}^{\rm ex}
(\alpha)
\end{equation}

\subsubsection{Expression for the structure of the fluid in terms
	       of the excess chemical potential of the components}

In order to relate the structure of the fluid to the
excess chemical potential of the components of the mixture,
(i.e., $A$, $B$, $C$, etc. monomers and n-mers), let us consider
the thermodynamic cycle of figure 1.

In the first step of the cycle, an isolated n-mer is dissolved
into a fluid mixture with association degree $\alpha$.
 Initially, the total Gibbs free energy
of the system is the sum of the free energy of the isolated n-mer,
$G_{\rm n-mer}$ and the free energy of the mixture, $G_{\rm mix}(\alpha)$. 
After dissolving the n-mer, the resulting free energy is that of the
original mixture with an extra n-mer, $G_{\rm mix+n-mer}$. The change in $G$ 
is therefore:
\begin{equation}
\Delta G_1 = G_{\rm mix+n-mer} - G_{\rm mix} - G_{\rm n-mer} 
\end{equation}
In the thermodynamic limit, the difference $G_{\rm mix+n-mer} - G_{\rm mix}$ 
becomes equal to the chemical potential of the n-mer in the mixture,
while $G_{\rm n-mer}$ may be considered to be the chemical potential
of the isolated n-mer (i.e., the free energy difference between
a system with a single n-mer and an empty system). It is thus
seen that:
\begin{equation}
\label{eqDG1}
\Delta G_1 = \mu_{\rm n-mer}^{\rm ex}(\alpha)
\end{equation}

In a second step, $n$ uncorrelated monomers of type $A$, $B$, $C$, etc. 
dissolved in the mixture
are brought together in such a way that the resulting $ABC\cdots$ complex
forms one of the many possible conformers of the n-mer, say, one
such that the vector joining $B$ to $A$ is ${\bf{r}}_{12}$, that joining
$C$ to $B$ is ${\bf{r}}_{23}$, etc.
This event will occur according to a probability density
given by the n-body correlation function of the mixture, 
$\rho^n g({\bf{r}}_{12},{\bf{r}}_{23},\ldots,{\bf{r}}_{n-1,n})$.
The n-mer density is given as an integral of this function
over all the conformations compatible with the monomer:
\begin{equation} \label{eqdensity}
\rho_{\rm n-mer}  = \rho^n \int_v \cdots \int_v
	         g({\bf{r}}_{12},{\bf{r}}_{23},...,{\bf{r}}_{n-1,n}) 
                {\rm d}^3 {\bf{r}}_{12} {\rm d}^3{\bf{r}}_{23} \cdots {\rm d}^3 {\bf{r}}_{n-1,n}
\end{equation}
where $v$ is the volume within the range of the bonding potential.
i.e., any two adjacent monomers whose distance vector is not
within this volume are not considered to be bonded.

The process of forming the n-mer from a set of n uncorrelated
monomers may be considered as a chemical reaction of the form
\begin{center}
\begin{tabular}{ccc}
 n uncorrelated monomers & $\rightleftharpoons$ & n correlated monomers
\end{tabular}
\end{center}
Friedman \cite{friedman85} has shown that such an equation is characterized by 
an equilibrium constant of the form $K_{\rm eq} = \rho_{\rm n-mer}/\rho^n$.
Substitution of Eq. \ref{eqdensity} into the expression for the equilibrium
constant, yields $K_{\rm eq} = \rho^n \Delta/\rho^n$, where,
\begin{equation}
\label{eqDelta}
\Delta  =       \int_v \cdots \int_v
	         g({\bf{r}}_{12},{\bf{r}}_{23},...,{\bf{r}}_{n-1,n}) 
                {\rm d}^3{\bf{r}}_{12} {\rm d}^3{\bf{r}}_{23} \cdots {\rm d}^3{\bf{r}}_{n-1,n}
\end{equation}

Now, the free energy of the whole process may be written down as:
\begin{equation}
\Delta G = \Delta G_2 + kT\ln K_{\rm eq}
\end{equation}
However, when the system reaches equilibrium, $\Delta G = 0$, so 
that we finally obtain:
\begin{equation}
\label{eqDG2}
\Delta G_2 = - kT \ln \Delta
\end{equation}

Similar arguments as those put through for the first and 
second steps of the cycle, lead to the conclusion that 
\begin{equation}
\label{eqDG3}
\Delta G_3 = \sum_{i=1}^{n} \mu_i^{\rm ex}(\alpha)
\end{equation}
\begin{equation}
\label{eqDG4}
\Delta G_4 = -kT \ln \Delta_0
\end{equation}
where $\Delta_0$ is the integral of Eq. \ref{eqDelta} evaluated at zero
density.

Substitution of Eq. \ref{eqDG1}, \ref{eqDG2}, \ref{eqDG3}, \ref{eqDG4}
into the net energy balance of the cycle, 
$\Delta G_1 + \Delta G_4 =  \Delta G_3 + \Delta G_2$,
leads to the desired equation relating the structure of the fluid with 
the excess chemical potential of the components of the mixture:
\begin{equation}
\label{eqDeltatermo}
kT \ln \frac{\Delta}{\Delta_0} = \sum_{i=1}^{n} \mu_i^{\rm ex}(\alpha) -
				               \mu_{\rm n-mer}^{\rm ex}(\alpha)
\end{equation}
A similar equation has been derived in a rather more formal way recently
\cite{smith98} for the particular case of infinitely short ranged 
association potentials. The derivation we have employed is based on the
thermodynamic cycle presented by Zhou and Stell \cite{zhou92}, which
allows to extend their previous result to finite range potentials.

\subsubsection{Expression for the degree of association in terms
	       of the structure of the fluid}

Substitution of Eq. \ref{eqalfatermo} into Eq. \ref{eqDeltatermo} 
shows that the equilibrium constant is related to
the n-body correlation function of the associating mixture
through the following relation:
\begin{equation}
\frac{K}{K_0} = \frac{\Delta}{\Delta_0}
\end{equation}

Finally, using the expression for $K$ in terms of the degree
of association, Eq. \ref{eqKeq}, we obtain the desired equation relating
the degree of association with the structure of the system:
\begin{equation}
\label{eqalfa}
  \frac{ \alpha }{ (1-\alpha)^n\rho^{n-1} } = \frac{K_0}{\Delta_0} \Delta
\end{equation}

\subsection{The equation of state}

Previously, we obtained an approximate equation for the pressure
in terms of the association degree of the mixture, Eq. \ref{eqP_1}.
The density derivative of the association degree, required in
such an expression is obtained from Eq. \ref{eqalfa}:
\begin{equation}
\label{eqalfader}
\frac{\partial \alpha}{\partial \rho} = \frac{1-\alpha}{\rho} \cdot
 \frac{(n-1)\alpha + \rho \alpha \frac{ \partial \ln \Delta}{\partial \rho} }
         {1 + \alpha (n-1) }
\end{equation}
Substitution of this result into Eq. \ref{eqP_1} yields an expression
for the change in pressure due to the formation of the n-mers
from a fluid of non-associated monomers. This expression depends solely on the
degree of association and the n-body correlation function of
the associating system:
\begin{equation}
\label{eqP_2}
\frac{ \Delta p}{kT \rho} = - \alpha \lgroup n - 1 + \rho 
           \frac{\partial \ln \Delta}{\partial \rho} \rgroup
\end{equation}

By taking the limit of infinite association, which physically corresponds
to infinitely increasing the well depth of the bonding potential,
we would arrive at an equation for the pressure of the n-mer fluid
relative to that of the monomer reference fluid.
In order to do so, however, we would require the
n-body correlation function of the associating system, which
enters through $\Delta$. Unfortunately, quantitative understanding of such
 high order 
correlation
functions is far beyond our present knowledge.
We will therefore need to make some further approximations 
in order to get a tractable expression for the pressure.

\subsubsection{Decoupling of the n-body correlations}

In order to simplify the problem of the n-body correlations,
we invoke a so called 'linear' decoupling approximation,
which attempts to describe the n-body correlation function
in terms of $n-1$ two body correlation functions:
\begin{equation}
g^{(n)}({\bf{r}}_{12},{\bf{r}}_{23},\cdots,{\bf{r}}_{n-1,n}) =
g^{(2)}({\bf{r}}_{12})g^{(2)}({\bf{r}}_{23})\cdots g^{(2)}({\bf{r}}_{n-1,n})
\end{equation}
Here it should be understood that $g^{(2)}({\bf{r}}_{12})$ stands
for the pair correlation function of monomers of type $A$
with monomers of type $B$ in the multicomponent mixture
of associating monomers, $g^{(2)}({\bf{r}}_{23})$ stands for
the pair correlation function of monomers of type $B$ with
monomers of type $C$ and so on.
Still, these two body correlation functions 
may be quite difficult to obtain.
In order to simplify the problem, consider one of these
correlation functions, say, $g^{(2)}({\bf{r}}_{12})$,
in the limit of zero density:
\begin{equation}
g^{(2)}({\bf{r}}_{12};\rho=0) = \exp(- [u_0({\bf{r}}_{12}) + \Phi({\bf{r}}_{12})]/kT )
\end{equation}
It can be seen that, in this limit, the pair correlation function may
be exactly expressed in terms of the pair correlation function
of a reference system with no bonding potential,
$g_0^{(2)}({\bf{r}}_{12})$, times the Boltzmann factor
of the bonding potential:
\begin{equation}
g^{(2)}({\bf{r}}_{12};\rho=0) = g_0^{(2)}({\bf{r}}_{12};\rho=0)
	                   \exp(- \Phi({\bf{r}}_{12})/kT )
\end{equation}
Using the linear approximation to the n-body correlations and considering 
the above equation to hold true at any
density, the $\Delta$ integral is simplified considerably,
giving:
\begin{equation}
\label{eqDeltan}
\Delta = \delta^{n-1}
\end{equation}
where $\delta$ is defined as:
\begin{equation}
\label{eqdelta}
\delta =  
 \int_v g_0^{(2)}({\bf{r}}_{12}) \exp(- \Phi({\bf{r}}_{12})/kT ) {\rm d}^3{{\bf r}}_{12}
\end{equation}

\subsubsection{An equation of state for the n-mer in terms of the
               thermodynamics and structure of the monomers}

As a consequence of the two approximations given for the n-body correlations,
we are now able to write down an expression for the pressure
of the n-mer in terms of the properties of the reference fluid. 
Indeed, after setting $\alpha=1$, substitution of Eq. \ref{eqDeltan} into 
Eq. \ref{eqP_2} 
leads finally to  the following result:
\begin{equation}
\label{eqP_3}
\frac{\Delta p}{kT \rho} = - [ n - 1 ][1 + \rho 
           \frac{\partial \ln \delta}{\partial \rho} ]
\end{equation}
Note that as we are considering the limit of complete
association,  $\rho$ is equal to the polymer number density.

Adding the contribution of the reference system to the previous equation,
we are now able to
write down an equation for the compressibility factor,
$Z= p/kT \rho$ of the fluid of n-mers:
\begin{equation}
\label{eqZ}
Z_{\rm n-mer} = n Z_0 - (n-1) [1 + \rho \frac{\partial \ln \delta}
							   {\partial \rho} ] 
\end{equation} 
where $Z_0$ is the compressibility factor of the reference fluid,
measured at the same monomer density as the n-mer fluid.
This is a rather remarkable equation, as it gives the equation
of state of a chain fluid from the properties of a fluid
of monomers alone. Different versions of this equation will arise
from the different theories available to describe the structure
and thermodynamics of the fluid of monomers. In section IV
we will consider two such theories in order to describe our model
polymer. Let us recall at this point, however, that a simple, qualitative 
version
of Eq.\ref{eqZ} may be obtained by simply considering that
$\delta$ does not depend on the density. In this way, the resulting
equation does no longer depend on the structure of the reference fluid.
Nikitin {\em et al.}\cite{nikitin96} have explored this equation using the van der Waals
equation of state for $Z_0$ and find the same qualitative behavior as is
found in this work.

\subsection{Comparison with Wertheim's theory of association}

The arguments we have put through in order to arrive at 
Eq. \ref{eqZ} are rather physical and  intuitive.
On the other hand, Wertheim has developed a very general
theory of association based on  a re-summed cluster expansion,
where the significance of each of the approximations is
mathematically well understood. It is interesting
to compare the results of this rather formal theory with
the physically appealing description that we have used,
largely based on the work of Zhou and Stell \cite{zhou92}.

In the extension of Wertheim's theory of association,
the compressibility factor of the chain molecule is given
as:
\begin{equation}
\label{eqZw}
Z_{\rm n-mer} = n Z_0 - (n-1) [1 + \rho \frac{\partial \ln \kappa}
							   {\partial \rho} ]
\end{equation}
where $\kappa$ is defined as:
\begin{equation}
\kappa =  
 \int_v g_0^{(2)}({\bf{r}}_{12}) [\exp(- \Phi({\bf{r}}_{12})/kT) -1 ] 
                                     {\rm d}^3{\bf r}_{12}
\end{equation}

The limit of complete association requires that $\Phi$ have an 
infinite well depth, so, within most of the range of the bonding
potential, the Boltzmann factor is exceedingly bigger than 
$1$. Therefore,  $\kappa$ and $\delta$ become identical and
Eq. \ref{eqZw} is essentially equal to Eq. \ref{eqZ}.

Before proceeding to the next section, let
us first summarize the approximations
invoked to obtain Eq. \ref{eqZ}: 

\begin{enumerate}
\item Assume that the free energy difference between the reference
      system and the completely associated system
      takes the form of the low density limit in all the density range.
\item Decouple the n-body correlation function of the associating system into
       $n-1$ two body
      correlation functions through a linear approximation.
\item Assume that the two body correlation function of the associating
      system is given in terms of the two body correlation function of
      a reference system with no bonding potential, as suggested by the 
      exact low
      density limit.
\end{enumerate}

\section{Predictions for the scaling laws of the critical properties}

We start by assuming that the critical density does become small
for large chain lengths, so that one can describe the equation of
state in terms of a truncated virial expansion.
\begin{equation} \label{eqvirseries}
\frac{p}{kT} = \rho + B_2(T) \rho^2 + B_3(T) \rho^3
\end{equation}
where $\rho$ is the polymer number density.
By applying the conditions for the critical point of pure fluids, i.e.,
\begin{equation}
\begin{array}{c}
\left( \frac{\partial p}{\partial V} \right)_{T_c} = 0    \\
\left( \frac{\partial^2 p}{\partial V^2} \right)_{T_c} = 0 
\end{array}
\end{equation}
we obtain a set of equations for the critical temperature
and density:\cite{vega96a}
\begin{eqnarray}
\label{sistema}
B_2(T_c) +  \sqrt{3B_3(T_c) } &=& 0  \\ \label{eqrhoc}
\sqrt{ 3 B_3(T_c)  }\rho_c &=&  1
\end{eqnarray}

By making a Taylor expansion on powers of the density, the
first and second virial coefficients predicted by Wertheim's
equation are found to be:
\begin{equation}
\begin{array}{c}
B_2 = n^2 \left( b_2 - \frac{n-1}{n} a_2 \right)  \\
B_3 = n^3 \left( b_3 - \frac{n-1}{n} a_3 \right)
\end{array}
\end{equation}
where $b_2$ and $b_3$ are the second and third virial
coefficients of the reference fluid of non--bonded monomers,
while $a_2$ and $a_3$ are the zeroth and first order
coefficients in a monomer density expansion of 
$ \partial \ln \delta/\partial \rho$.
Of course, all these quantities are chain length independent.

Now, in order to solve Eq. \ref{sistema} for
the critical temperature we will need to linearize
the virial coefficients with respect to the temperature.
To do so, let us assume for the time being that there is
a finite asymptotic critical temperature in the limit
of infinite chain length, which we call $\Theta$, in
analogy with the polymer + solvent case. 
We now make a series expansion of $B_2$ and $B_3$ 
in powers of $\Delta T = \Theta - T$ up to first order, and consider the limit
of this expression for large $n$, leading to 
\begin{equation}
\begin{array}{l}
B_2(T) = n^2 \left( C_2 - C_2' \Delta T \right) \\
B_3(T) = n^3 \left( C_3 - C_3' \Delta T \right) 
\end{array}
\end{equation}
where $C_2=b_2(\Theta)-a_2(\Theta)$ and $C_3=b_3(\Theta)-a_3(\Theta)$ while    
$C_2'$ and $C_3'$ are the corresponding derivatives with respect to temperature.
Substitution of the linearized virial coefficients into the
condition for the critical temperature leads to a quadratic
equation for $\Delta T$. Solving for this equation yields $\Delta T_c(n)$,
defined as $\Theta - T_c(n)$:
\begin{equation} \label{eq:quadratic}
\Delta T_c(n) - \frac{C_2}{C_2'} =
           \pm \frac{1}{2C_2^{' 2}} \left( 12C_2^{' 2}C_3 - 12C_2C_2'C_3' +
           9C_3^{' 2} \frac{1}{n} \right) ^{1/2} \;\; \frac{1}{n^{1/2}}
             - \frac{3C_3'}{2C_2^{' 2}} \;\; \frac{1}{n}
\end{equation}
This equation shows that $\Delta T_c(n)$ must reach an asymptotic finite
value, since the right hand side term should ultimately vanish for large
$n$. The requirement for $T_c(n)$ to attain a finite asymptotic critical
temperature equal to $\Theta$ is then obeyed provided that $C_2$ vanishes.
If we now notice that
\begin{equation}
\lim_{n \to \infty} B_2(\Theta) = n^2 C_2
\end{equation}
we arrive at the conclusion that indeed $C_2$ must vanish at
the Boyle temperature of the infinitely long polymer, $T_B^{\infty}$,
thus identifying the $\Theta$ temperature with the Boyle temperature
of the infinitely long polymer.
From the definition of $C_2$ we see that this temperature is
attained when the following condition is obeyed:
\begin{equation}
\label{eqtheta}
b_2(T_B^{\infty}) - a_2(T_B^{\infty}) = 0
\end{equation}
Note also that the leading terms of the expansion (Eq.
\ref{eq:quadratic}) are of order $n^{-1/2}$ and $n^{-1}$, just as predicted 
by the FH theory.

The case of the critical polymer density is much simpler. Substitution
of the expression for $B_3$ in the condition for the critical
density shows that:
\begin{equation}
\rho_c(n) \propto n^{-3/2}
\end{equation}
so that the critical mass density decreases with a power
law proportional to $n^{-1/2}$, as predicted by the FH theory.

It is important to note that the above arguments apply regardless
of the specific form in which the reference fluid 
(thermodynamics and structure) is described. In particular,
the simplest implementation of TPT1, proposed by Nikitin {\em et al.}
\cite{nikitin96} considers $\delta$ to be a constant. In such a 
case, $a_2$ is zero at all temperatures. However, Eq. \ref{eqtheta} shows
that this simple version still predicts an asymptotic critical
temperature which must obey the condition $b_2(T_B^{\infty})=0$.
Obviously, this condition is obeyed for the Boyle temperature
of the reference fluid.

Another interesting issue is the apparent universality of the
compressibility factor as predicted by the truncated virial expansion of
Eq. \ref{eqvirseries}. Indeed, substitution of this equation into the 
condition for
the critical point shows that, apart from Eq. \ref{eqrhoc} it must also hold
that $\rho_c = - B_2/(3B_3)$. Using this expression for the critical
density in the linear term of Eq. \ref{eqvirseries} and Eq. \ref{eqrhoc} in 
the quadratic
term, it is seen that both terms cancel each other exactly. 
Dividing the resulting expression for the pressure by the critical
density (Eq. \ref{eqrhoc}) then shows that:
\begin{equation} 
Z_c(n) = \frac{p_c}{\rho_c kT_c} = \frac{1}{3} + 
                  \frac{B_4}{(3B_3)^{3/2}} + \ldots
\end{equation} 
Obviously, this result is independent on whatever assumption is made
concerning the actual $n$ dependence of the virial coefficients and
shows that a finite asymptotic critical compressibility factor of about $1/3$
is expected in the limit of infinite chain length, irrespective of the
nature of the polymer. In the context of TPT1,  a constant compressibility
factor implies that the critical pressure must decrease as $n^{-3/2}$.

\section{Application to a polymer model}

Let us consider a polymer model as that described in the previous
section, with the reference fluid considered to be
a truncated and shifted potential of the form:
\begin{equation}
u_{0}(r)= \left\{ \begin{array}{ll}
V_{LJ}(r) - V_{LJ}(r_c) & r \leq r_c \\
0 & r > r_c
\end{array}   \right.
\end{equation}

where $r_c=2 \cdot 2^{1/6}$ and $V_{LJ}$ is the usual 
Lennard-Jones potential,

\begin{equation}
V_{LJ}(r) = 4 \epsilon \left\{ \left(\frac{\sigma}{r}\right)^{12} -
                                \left(\frac{\sigma}{r}\right)^{6} \right\}
\end{equation}

As to the bonding potential responsible for the connectivity between
adjacent monomers, we will consider the FENE potential, defined
in terms of $R_0$, the maximum displacement between monomers
and $k_0$, a sort of elastic constant:
\begin{equation}
\Phi(r)  =  \left\{ \begin{array}{ll}
 - k_0 R_0 ^2 \ln ( 1 - \frac{r^2}{R_0 ^2} )    - E_b   &  0<r<R_0 \\
  0                                                &   r \ge R_0 
\end{array}   \right.
\end{equation}
In order to ensure permanent connectivity of the n-mer,
a constant, $E_b$, which is (conceptually) made infinitely large,
is added to the actual FENE potential.

In what follows, we will set $k_0 = 15 \epsilon / \sigma^2$
and $R_0 = 1.5 \sigma$ and use the Lennard-Jones energy and
range parameters as energy and length units, respectively.
At liquid--like densities the most probable distance between
non--bonded monomers is about $1.12 \sigma$, which is bigger than the
most probable distance $0.96 \sigma$ between bonded monomers.
Note at this point that most of the previous applications of
Wertheim's theory have been restricted to bonding potentials
of infinitely short range, allowing for a single possible
bond length. To our knowledge, only once has the effect of
a soft bonding potential been considered previously. \cite{ghonasgi94}

In section II we related the equation of state of such a polymer fluid 
to the properties of the reference system of LJ monomers. 
What is now required is a theory for both the thermodynamics
and the structure of the monomer fluid. We have obtained the required 
input from integral equation theory and thermodynamic perturbation 
theory. Let us consider each of them in turn.

\subsection{The RHNC integral equation theory}

In this approach, one attempts to calculate the exact pair
correlation function, which is then used to
evaluate the mechanical properties of the fluid. 
The Ornstein-Zernike equation relates the total pair correlation function,
$h(r)=g(r)-1$ to a short range direct correlation function $c(r)$:

\begin{equation}
\label{oz}
h(r_{12}) = c(r_{12}) + \rho \int h(r_{13}) c(r_{23}) {\rm d}^3 {\bf r} _{3}
\end{equation}

Additionally, this integral equation must be provided with a closure 
that relates
$c(r)$ to $h(r)$. We use the Reference Hyper-netted chain equation of Lado and
Ashcroft. \cite{lado73,lado83} 
Although this set of equations can only be solved numerically
and convergence is not a trivial matter, an efficient algorithm
due to Labik and Malijevsky \cite{labik85} makes the calculations
affordable with a modest amount of CPU time.
Once $g(r)$ is known, the pressure of the fluid may be calculated
using the standard relation \cite{mcquarrie76}:
\begin{equation}
\label{eqvirial}
\frac{p}{k_BT} = \rho - \frac{\rho^2}{6k_BT} \int r \frac{{\rm d}u_0}{{\rm d}r}\;\;
								g(r) \;\;4 \pi r^2 {\rm d}r
\end{equation}
Similarly, $\delta$ may be calculated using Eq.\ \ref{eqdelta}.
Actually, solving the integral equation for each of the desired
thermodynamic states may result rather cumbersome. In practice
we solve the OZ+RHNC equation for several hundreds of state points
and fit the pressure and $\delta$. Details of the procedure may
be found in Appendix A.

\subsection{The perturbation theory of Tang and Lu}

Perturbation theory was the first approach to give quantitative
results for the thermodynamics of simple fluids at high density.
\cite{barker67,weeks71} Compared to integral equation theory, it
gives similar results at high densities at a smaller computational
cost, with the advantage that the free energy is obtained directly,
without the need for thermodynamic integration. On the other hand, 
the traditional
perturbation theories of Barker and Henderson \cite{barker67} and
Weeks-Chandler-Andersen \cite{weeks71} are known to be rather poor
at low densities because the underlying assumptions of these
theories no longer hold true. \cite{tang97c}
Fortunately, Tang and Lu have presented rather recently a second order
perturbation theory for Lennard-Jones fluids which is very
accurate both at low and high densities. \cite{tang97a,tang97b}
The success of this theory relies on a rather good description
of the structure of the fluid, which is obtained from the
OZ equation, supplemented by a simple closure known as the Mean Spherical 
Approximation. The use of this closure is very convenient
because it has allowed to obtain a very good approximation to
the actual free energy with a purely analytical equation.
Furthermore, Tang and Lu have been able to obtain also analytic
expressions for the pair correlation function using the MSA
closure. \cite{tang93,tang97c}
With minor modifications we were able to extend this theory to
our truncated and shifted Lennard-Jones potential and obtain
a lengthy but analytic expression for the thermodynamics and structure of
our reference fluid. Details of the implementation are explained
in Appendix B.

In what follows, we shall present the results of Wertheim's theory
for our polymer model using both the RHNC and the MSA thermodynamic
theories for the reference fluid. We will call each of the versions
TPT1-RHNC and TPT1-MSA, following the original name for 
Eq. \ref{eqZ} due
to Wertheim. \cite{wertheim87}

\section{Simulation details}

In order to test the TPT1 theory, we have performed extensive computer
simulations. Chain length $n=10$ was chosen for a detailed comparison
to the theory.  We have calculated the pressure and chemical potential for 
5 temperatures $T=1.68, 2.5, 3.0, 4.0$, and 5. The lowest value corresponds 
to a 
subcritical isotherm, while the highest value is above the $\Theta$ temperature.
The pressure isotherms were evaluated from the virial in 
NVT Monte Carlo simulations
The length $L$ of the (cubic) simulation box was fixed to 
18 $\sigma$ units.
The density dependence of the chemical potential was calculated using
grand-canonical Monte Carlo simulations. Chain conformations were sampled
using local monomer displacements and slithering snake like motions. Particle 
insertions and  deletions were performed following configurational bias grand-canonical 
acceptance rules. \cite{wilding96,smit95b,frenkel96} 
In order to obtain the equation of state we employ  cycles
of 25 local moves, 25 reptations and 10 CBGC moves. About 40000 such 
cycles were performed so that at least a few thousand particle 
insertion-deletion attempts were accepted. The volume of the simulation box 
was chosen so that an average number of about 50 chain molecules was obtained.


In order to compare the theory to the simulations, we must make sure
that the chemical potentials are expressed with respect to the same 
reference state.
In order to do so, we define the excess chemical potential to be the 
difference between $\mu$ and the chemical potential of an ideal gas of chains 
$\mu_{\rm id}$ with the full intramolecular interactions but no intermolecular 
interactions. 
\begin{equation}
\mu^{\rm ex} \equiv \mu - \mu_{\rm id} \qquad \mbox{with} \qquad \mu_{\rm id} = \ln \rho - (n-1) \ln C - \ln \langle W_0 \rangle
\end{equation}
The first term denotes the translational entropy. Contributions due to the
integration over the momenta are ignored, because they contribute equally to
the reference system and the interacting polymer liquid. To determine the
ideal gas contribution we construct chains according to the Rosenbluth 
procedure.  The distance $l$ between bonded neighbors is chosen according 
to its Boltzmann weight $p(|l|) = 4 \pi l^2 \exp( - (u_0 + \Phi)/kT )/C$ 
where $C$ is the normalization constant.  $W_0$ denotes the Rosenbluth 
weight of the chains due to non--bonded interactions, measured at zero
density. Once the excess chemical potential has been obtained, it is
compared to the excess chemical potential as predicted by the
theory, which is evaluated using the standard thermodynamic relation:
\begin{equation}
 \frac{\mu^{\rm ex}}{k_BT} = \frac{A^{ex}}{Nk_BT} + Z -1
\end{equation}

The grand-canonical ensemble allows also for an accurate measurement of
the phase diagram, because the order parameter (i.e., the density)
is not conserved and density fluctuations are efficiently equilibrated.
We monitor the probability distribution $P(\rho)$ of the density. Close to 
two phase coexistence, the probability distribution is bimodal: one peak 
corresponds to the vapor, the other corresponds to the liquid. The coexistence
chemical potential $\mu_{\rm coex}$ is fixed by the condition of equal weight in both peaks:\cite{mw94}
\begin{equation}
\int_0^{\rho^*} {\rm d}\rho\; P(\rho) \stackrel{!}{=} \int_{\rho^*}^\infty {\rm d}\rho \;P(\rho)
\qquad
\mbox{with}
\qquad
\rho^* = \int_0^\infty {\rm d}\rho\; \rho P(\rho)
\end{equation}
Far below the critical points the probability between the two peaks is
very low, and we use a re-weighting scheme as to encourage the system to
``tunnel'' between the two phases. To this end we add a term $k_BT \ln W(\rho)$ to 
the original Hamiltonian. Choosing $W(\rho) \approx P(\rho)$ the system visits
all densities with roughly equal probability. The probability distribution of the
grand-canonical ensemble is obtained via re-weighting the distribution in the
simulations $P_{\rm MC}$ according to $P(\rho)=P_{\rm MC}(\rho)W(\rho)$.
At very low temperatures the density of the liquid in coexistence with
its vapor becomes very high and the configurational bias scheme becomes quite
inefficient. Since the density of the vapor is very low, however, its pressure is vanishingly
small. Hence, we employed NpT simulations at zero pressure to obtain the liquid
density at coexistence.

At the critical point the correlation length of density fluctuations
diverges and universal behavior is expected. For finite chain length the 
unmixing transition exhibits
3D Ising universal behavior. We have located the critical point for chain 
length $n=1, 10, 20, 40,$ and $60$ by mapping the symmetrized order parameter 
distribution $P_{\rm sym}(\rho)=[P(\rho)+P(\rho_c-\rho)]/2$ onto the universal 
scaling function of the 3D Ising model. This symmetrization reduces field mixing
corrections which are antisymmetric in $\rho - \rho_c$ to leading order.
Normalizing $P_{\rm sym}$ to unit variance and norm we eliminates all 
non--universal factors. The results of this mapping are presented in 
Fig.\ref{fig:mapping}, where we have used system sizes 
$L=11.3, 13.8, 18, 22.5,$ and $27$
for chain length $n=1, 10, 20, 40$, and $60$, respectively. This method gives 
an accurate 
location of the critical temperature and density (finite size corrections 
to the critical 
density of the order $L^{-(1-\alpha)/\nu}$ are neglected\cite{mw94}). The locations of 
the critical points are collected in Tab. \ref{tab:cripoint}.

\section{Results and Discussion}

Let us first examine the thermodynamic data for the chains of
10 monomers.
Fig.\ref{fig:pressure} shows the predictions of TPT1 for several pressure 
isotherms ($kT/\epsilon$=5, 4, 3, 2.5 and 1.68) compared 
with simulation results. Both the RHNC and MSA versions of the
theory are seen to give rather good estimates; at
the highest temperatures, far above the estimated $\Theta$ point
of our model (see below) as well as at the lowest, 
a subcritical isotherm. Overall, the RHNC version seems to
describe the isotherms slightly better.  

Results for the excess chemical potential of the chains are shown
in Fig.\ref{fig:chempot}. The agreement is also quite satisfactory, though the
results are slightly worse than for the pressure isotherms, 
specially at the lowest temperatures and densities. 
Indeed, the main assumption of the theory, that the
local environment of a monomer in the polymer fluid is similar to
that of the monomer fluid breaks down in the low density limit.
The fluid is then made of isolated clusters of $n$ monomers,
rather than of single monomers uniformly distributed in space.
Likewise, the theory is unable to describe the density dependence of the single chain
internal energy and entropy.

The liquid--vapor coexistence curve of the 10-mer as obtained
from simulation and theory is shown in Fig.\ref{fig:phasen}. Both the RHNC
and MSA versions overestimate the critical temperature as
obtained from simulation by about $15\%$. Of course, this is expected for any classical 
theory. On the other hand,  far away from the
critical point, results from both versions of the theory
are seen to yield fair agreement with simulation. The MSA
version is somewhat more convenient, however, because it 
allows to calculate the coexistence at low temperatures with
no additional cost, while it becomes rather problematic to calculate
the coexistence for the RHNC version below the reference fluid
critical temperature. The reason for this is that the RHNC
integral equation presents a region of no solutions below this
point, so that the resulting equation of state is no longer defined
inside the liquid-vapor envelope.

We have also investigated the critical points of chains of 20, 40
and 60 n-mers, in an attempt to study the behavior of TPT1 for longer 
chains. Table \ref{tab:cripoint} gives a summary of the simulation results, 
obtained
by finite size scaling, together with predictions from TPT1-RHNC and
TPT1-MSA. Both versions overestimate the critical temperatures
by about 15\% for all chain lengths studied.
However,
the MSA and RHNC predictions seem to converge as the chain
length increases. On the other hand, the critical monomer 
densities are always underestimated, though the MSA version
seems to give much better agreement than the RHNC version.
In the latter theory the density decreases much too fast compared to the MC results. The overall behavior of the
critical parameters is illustrated in Fig. \ref{fig:critical point},
where both $T_c$ and $\rho_c$ are plotted against $n^{-1/2}$, the
predicted asymptotic scaling law for both of these properties. It is
seen that for chain lengths up to 60 monomers, the critical properties
are far from reaching their asymptotic behavior, so that the
simulations
do not allow as to asses unambiguously the predicted scaling laws.

Although the calculation of the critical point of fluids larger
than about 100 monomers by computer simulation becomes prohibitively
expensive, we can estimate the $\Theta$ point of  our polymer model
by an analysis of the temperature dependence of the polymer extension. Fig. \ref{fig:theta} shows a plot of the mean squared end to
end distance divided by $n-1$ as a function of temperature for
various chain lengths. In the infinite chain length limit, 
the intercept of two such plots occurs at the $\Theta$ point of the
polymer model. Extrapolation of the results gives as an 
estimate $\Theta \approx 3.3$.
As to the theory, fitting the critical temperature predicted
by TPT1-RHNC to a power law of the form 
$T_c=T_c^{\infty} + b n^{-1/2} + c n^{-1}$ in the range $10^2$ to
$10^7$ gives $T_c^{\infty}=3.44$. On the other hand, by searching for the
root in Eq. \ref{eqtheta}, we find that TPT1-MSA predicts
$T_c^{\infty} = 3.14$.
Assuming that the $\Theta$ point is indeed the critical point of
the infinitely long chain, as suggested by the considerations of 
Section III, it would seem that TPT1 is capable of giving an
excellent prediction for the $\Theta$ point of the polymer, even though
the actual prediction may vary somewhat depending on the theory
used to describe the reference fluid.
Remarkably, considering that $\delta$ is density independent
and using the simple van der Waals equation of state, Nikitin et
al. \cite{nikitin97}
have shown that TPT1 predicts an asymptotic critical temperature
$T_c^{\infty}=\frac{27}{8}T_c^{0}$. As $T_c^{0}$, the critical
temperature of the reference fluid, is approximately $1.0$ LJ
reduced units,
we find that the simplest TPT1 approach
already gives an excellent prediction for the $\Theta$ point of
$T_c^{\infty}=3.375$. This is, however, somewhat fortuitous as 
it was shown in Section III that
this TPT1-van-der-Waals approach of Nikitin actually predicts
that $\Theta$ is equal to the Boyle temperature of the
monomer fluid, $T_B^{0}$, which is about 
$T_B^{0}=2.58$. 
Thus, the relatively good estimate of $T_c^{\infty}$ turns out to
be a consequence of the over prediction of $T_B^{0}$
implicit in the van der Waals equation of state.

Recently, Chatterjee and Schweizer \cite{chatterjee98} have analyzed 
the behavior of
the critical point of infinite chain lengths using the PRISM theory.
For two of the closures employed, the same behavior as that predicted
by TPT1 is observed, at least concerning i) vanishing critical density
and ii) finite critical temperature. The power laws are, however,
different. The critical monomer density is predicted to vanish with a weaker
dependence which may be either $-1/3$ or $-1/4$ depending on whether
the RMPY/HTA or the MSA closures are used. The critical temperature
is predicted to reach a finite critical value with the same 
exponents as the critical density. 
However, there was no a priori reason for choosing one closure
over the others and several different trends could be obtained depending on the
closure that was used. It is pleasing to see that TPT1 is able to
give a unique conclusion, independent of the molecular theory
used to describe the monomer fluid.

\section{conclusion}

In this paper we have used the formalism of Zhou and Stell \cite{zhou92}
to extend the TPT1 theory to polymers with variable bond-length.
By using rigorous molecular theories for the reference fluid of
non--bonded monomers we have been able to explore two versions of the
theory that allow to  give a good description of the fluid 
without the need of any empirical data for the polymer.
Comparison with numerical simulations show that
both the RHNC version and the MSA version give good agreement
with the simulation data. At low temperatures, the RHNC version
seems to be more reliable, while at high temperatures there is
apparently little difference. The MSA version is seen to give
fair predictions for the critical points of the longer 
chains, with the advantage that it is almost analytic.
The $\Theta$ point of the model is predicted in very good agreement
by the RHNC version, as well as by the MSA version.

Concerning the critical behavior of long chains,
it has been shown that TPT1 predicts an approach of the critical
temperature to the $\Theta$ point with a power law of $n^{-1/2}$.
This is the correct behavior for the infinitely long polymer chain,
as mean field behavior must be recovered in the infinitely long
chain limit. The $\Theta$ point has been shown to be the
Boyle temperature of the infinitely long polymer while the
critical mass density is predicted to vanish with a power law
of $n^{-1/2}$. All of these predictions concerning the scaling
behavior of the critical points of the pure monomer fluid
are seen to agree exactly with the mean field (Flory--Huggins) predictions
for the critical behavior of polymer+solvent mixtures.
This gives further support to the idea that pure polymer equations
of state may be used as effective equations of state for the
polymer+continuum-solvent system and vice versa, polymer+solvent
equations of state may be used for the polymer+vacuum case;
a formal prove of this intuitively appealing idea is, however, difficult.

In a subsequent paper \cite{mueller99} we use the implementation of TPT1 
proposed here to describe the equation of state, together with
a self consistent field theory to study the surface and interfacial
properties of a polymer+solvent system.

\subsection*{Acknowledgments}

Helpful correspondence with Y.Tang is acknowledged.
The authors also benefited from stimulating discussions
with A. Milchev and V. Ivanov.
Financial support is due to grant Bi314/17 of the DFG and to
project No.PB97-0329 of the Spanish DGICYT. L.G.M. wishes to
thank the Universidad Complutense for the award of a predoctoral
grant and for founding a stay in Mainz.

\section{Appendix A: Fitting the RHNC data for the monomer fluid}

Rather than solving the integral equation of each of the state points
desired, which would be rather expensive, we followed a similar 
approach as that used by Johnson {\em et al.} \cite{johnson93} to describe the
Lennard-Jones fluid. We solved the RHNC integral equation for
a set of 756 states in the range $1.1\le T \le 6.0$ and 
$0 < \rho < 0.85$, we calculated the resulting compressibility factor
(see Eq. \ref{eqvirial}) and then fit the data to a 
Modified-Benedict-Webb-Rubin equation of state, given by:
\begin{equation}  \label{eq:bwr}
 \rho k T (Z - 1) = \sum_{i=1}^8 a_i\rho^{i+1}  +
                   F\sum_{i=1}^6 b_i\rho^{2i+1}
\end{equation}
where $F=\exp(-\gamma \rho^2)$, while the exact form of the $a_i$ and
$b_i$ coefficients is given in Table \ref{tab:coefbwr}. 
Once the fit to $Z$ is performed, the free energy may be determined from:
\begin{equation}
 A/N = \sum_{i=1}^8 \frac{ a_i\rho^{i} }{i}  +
       \sum_{i=1}^6 b_i G_i
\end{equation}
The $G_i$ coefficients obey the following recursive relation:
\begin{equation}
  G_i = - \frac{ F \rho^{2(i-1)} - 2(i-1)G_{i-1}}{2\gamma}
\end{equation}
with the first term given by $G_1=(1-F)/(2\gamma)$.
The parameters obtained for the fit are collected
in Table \ref{tab:parbwr}.

Contrary to the approach of Johnson {\em et at.},
we calculate the thermodynamics using the RHNC theory, rather than
computer simulations. In this way we are able to save several orders of
magnitude of CPU time. 
The effective diameter required in the RHNC
equation is determined as suggested by Lado {\em et al.}, \cite{lado83}
while the effective hard sphere bridge function is calculated from the 
parameterization of Labik and Malijevsky. \cite{malijevsky87,labik89}

Also required is the associating strength $\delta$ of the fluid.
This is obtained by solving equation \ref{eqdelta} for each of the 756 state
points. Rather than fitting $\delta$, which is a difficult task,
as it varies several orders of magnitude in the range $1.1\le T \le
6.0$, we fit the dimensionless ratio $\delta(\rho)/\delta(\rho=0)$.
The actual functional form used is:
\begin{equation}
\label{eqfit_delta}
\delta/\delta_0 = 1 + \sum_{i=1}^{5} \sum_{j=1}^{5} a_{ij} \rho^i
T^{(1-j)}
\end{equation}
The parameters for this fit are gathered in Table \ref{tab:pargr}.

\section{Appendix B: The perturbation theory of Tang and Lu}

In order to describe the thermodynamics of the reference fluid we
perform a Barker-Henderson decomposition of the monomer fluid pair
potential such that $u_0$ is described in terms of a repulsive reference
potential $w_{\rm ref}$, which is made of the positive region of $u_0$
and a perturbation, $w_{\rm per}$, which is made of the negative part of
the potential: \cite{barker67,mcquarrie76}
\begin{equation}
w_{\rm ref}(r) = \left \{ \begin{array}{ll} 
                  u_0(r)  & r \leq t\sigma  \\
                    0     & r  >   t\sigma  \\
                  \end{array} \right. \qquad \mbox{and} \qquad
w_{\rm per}(r) = u_0(r) - w_{\rm ref}(r) = \left \{ \begin{array}{ll}
                   0     &  r \leq t\sigma \\
                  u_0(r) &  r   >  t\sigma \\
                  \end{array} \right.
\end{equation}
where $t=1.0013$ defines the value of $r$ where $u_0$ becomes negative
(recall that we are considering a cut and shifted potential).
We now couple the perturbation potential to the reference potential
with a coupling parameter, $\lambda$, so that the actual potential
is recovered for $\lambda=1$:
\begin{equation}
w(r;\lambda) = w_{\rm ref}(r) + \lambda w_{\rm per}(r)
\end{equation}
At this stage we recall the fundamental functional expression that relates
the Helmholtz free energy with the radial distribution function,
\cite{hansen86}
\begin{equation}
\frac{\delta A}{\delta w(r;\lambda)} = \frac{1}{2}N\rho g(r;\lambda)
\end{equation}
Integration of this equation following the rules of functional calculus, 
\cite{hansen86}
leads to an expression relating the free energy of the monomer fluid
with that of the reference fluid:
\begin{equation}
\frac{A}{N} - \frac{A_{\rm ref}}{N} = \frac{1}{2} \rho
\int_{\lambda=0}^{\lambda=1} \int_{t \sigma}^{\infty} 
      g(r;\lambda) w_{\rm per} 4 \pi r^2 {\rm d} r {\rm d}\lambda
\end{equation}
It is then assumed that the radial distribution function may be
expanded as a series in powers of $\lambda$ of the form 
$g(r;\lambda)=g_{0}(r)+g_{1}(r)\lambda+...$. Truncation of the series 
to first order then yields:
\begin{equation}
\label{eqA2ndorderbis}
\frac{ A}{N}-\frac{ A_{\rm ref}}{N} = 
\frac{1}{2} \rho \int_{t \sigma}^{\infty} g_{0}(r)w_{\rm per}(r) 4 \pi r^2 {\rm d}  r
+
\frac{1}{4} \rho \int_{t \sigma}^{\infty} g_{1}(r)w_{\rm per}(r) 4 \pi r^2 {\rm d}  r
\end{equation}
Following Barker and Henderson \cite{barker67} we now choose to describe
the reference potential by means of a hard sphere fluid of appropriate
hard sphere diameter, $d$. This choice is justified because the 
reference potential is made essentially of the repulsive part of the
potential.  In this way, $\beta A_{\rm ref}/N$ and $g_0$ may be considered
to be the free energy and radial distribution function of an effective hard
sphere fluid, while $g_{1}$ may be solved 
using the MSA closure. \cite{tang97c} However, the integrals appearing
in the previous equation are still quite tedious to calculate and
a further approximation allows to get fully analytical results. This
is done by fitting the actual monomer potential, $u_0$ to a Two Yukawa
potential, following the procedure of ref. \cite{tang97a}:
\begin{equation}
\label{eqTYbis}
u(r)^{TY} = - k_0 \epsilon \frac{ e^{-z_1(r-t\sigma)} }{r}
            + k_0 \epsilon \frac{ e^{-z_2(r-t\sigma)} }{r}
\end{equation}
Actually, the fit needs to be performed only for values of $r$ greater
than $d$ and the resulting function (with $k_0=2.4405$, $z_1=3.492456$
and $z_2=13.109857$) is virtually identical to the
true potential. However, substitution of Eq. \ref{eqTYbis} in the first
and second order contributions of Eq. \ref{eqA2ndorderbis} and solving
$g_1$ for $u^{TY}$ rather than for $u_0$, a very accurate expression
for Eq. \ref{eqA2ndorderbis} may be obtained which is fully analytic.

This is done by rearranging Eq. \ref{eqA2ndorderbis} into the form:
\begin{equation}
\frac{\beta A}{N} = a_0 + a_1 +  a_2
\end{equation}
where $a_0$ is given by the Carnahan-Starling equation of state:
\begin{equation}
  a_0 = \frac{4\eta - 3\eta^2}{(1-\eta)^2}
\end{equation}
while
\begin{equation}
a_1 = \frac{1}{2}\rho 
          \int_{d}^{\infty} (g_0 - 1) w_{per}^{TY}  4\pi r^2 {\rm d} r +
      \frac{1}{2}\rho 
	    \int_{d}^{\infty} w_{per}  4\pi r^2 {\rm d}  r +
             g_0(d) \int_{t \sigma}^{d} w_{per}  4 \pi r^2 {\rm d} r
\end{equation}
and
\begin{equation}   
a_2 = \frac{1}{4} \rho 
          \int_{d}^{\infty} g_1(r)w_{per}^{TY} 4 \pi r^2 {\rm d} r +
      \frac{1}{4} \rho
      g_1(d) \int_{t \sigma}^{d} w_{per} 4 \pi r^2 {\rm d} r
\end{equation}
In the last two equations it is understood that $w_{\rm per}^{TY}$ is the 
perturbation potential
when expressed as in Eq. \ref{eqTYbis}; the actual $w_{\rm per}$ potential is
used where ever possible; $g_0=1$ and $g_1=0$ beyond the
cutoff distance of the potential while $g_0$ and $g_1$ are considered
to remain constant in the range $[d,t\sigma]$.
The only difference between the above expressions and those
obtained by Tang {\em et al.} \cite{tang97a,tang97b} for the true Lennard-Jones 
potential 
are found in the trivial integrals of the form $\int w_1 4 \pi r^2 {\rm d}  r$
because both the integration limits and the perturbation potential
differ.
Solving $a_1$ and $a_2$ yields:

\begin{eqnarray}
   a_1  & = & 
- \frac{12 \eta \beta \epsilon}{d^3} 
\bigg [ \; k_1 \left ( \frac{L(z_1 d)}
                           {z_1^2(1-\eta)^2 Q(z_1 d)} -
                       \frac{1+z_1 d}{z_1^2}
               \right)       
\nonumber \\ & &
\nonumber \\ & & \qquad \qquad
        -  k_2 \left ( \frac{L(z_2 d)}
	                     {z_2^2(1-\eta)^2 Q(z_2 d)} -
                       \frac{1+z_2 d}{z_2^2}
              \right)
\bigg ]
\nonumber \\ & &
\nonumber \\ & & 
  +  48 \eta \beta \epsilon W_{\rm cs}(r_c,d) 
  -  48 \eta \beta \epsilon g_0(d) W_{\rm cs}(t\sigma,d) 
\end{eqnarray}

\begin{eqnarray}
   a_2  & = &  
 - \frac{6 \eta \beta^2 \epsilon^2}{d^3} 
  \bigg [ \; \frac{k_1^2}{2z_1Q^4(z_1d)} +
             \frac{k_2^2}{2z_2Q^4(z_2d)} 
\nonumber \\ & &
\nonumber \\ & & \qquad \qquad \qquad
	    -  \frac{2k_1k_2}{(z_1+z_2)Q^2(z_1d)Q^2(z_2d)}
  \bigg ]
\nonumber \\ & &
\nonumber \\ & &
 - 24\eta \beta^2 \epsilon^2
  \left[   \frac{k_1/d}{Q^2(z_1 d)} - 
           \frac{k_2/d}{Q^2(z_2 d)}
  \right]
          W_{\rm cs}(t\sigma,d) 
\end{eqnarray}
where $\eta = \pi/6 \, d^3 \rho$ and 
$k_i = k_0 e^{z_i (t\sigma-d)}$. 
The Barker-Henderson diameter, defined as
\begin{equation} 
d = \int_0^{\infty} ( 1 - e^{-\beta w_{\rm ref}(r)} ) {\rm d}r
\end{equation} 
is parameterized using the formula proposed in Ref. \cite{desouza93}:
\begin{equation} 
d/\sigma = 2^{1/6} \left [ 1 +
                      \left ( 1 + \frac{T+c_2 T^2 + c_3 T^4}
			                     {        c_1        }
			    \right)^{1/2}
                   \right]^{-1/6}
\end{equation} 
where $c_1=1.150167$, $c_2=-0.046498$, and $c_3=0.0004477054$.

On the other hand, $Q$ is defined as:
\begin{equation}
   Q(t)  = \frac{S(t) + 12 \eta L(t) e^{-t}}{(1-\eta)^2 t^3}   
\end{equation}
where
\begin{equation} 
S(t) = (1-\eta)^2 t^3 + 6\eta (1-\eta )t^2 + 18\eta^2 t - 12 \eta(1+2 \eta) 
\end{equation} 
and
\begin{equation} 
L(t) = (1 + \eta / 2) t + 1 + 2\eta
\end{equation}

The reference radial distribution at contact is given by:
\begin{equation}
g_0(d) = \frac{1+\eta/2}{(1-\eta)^2}
\end{equation}
while $W_{cs}$ is to constant factors, the definite integral of the
perturbation potential, $w_{\rm per}$:
\begin{eqnarray} 
W_{cs}(x,y) & = & \Bigg \{  
        \frac{1}{3} \left [ \left( \frac{\sigma}{x} \right)^3  - 
	     	                \left( \frac{\sigma}{y} \right)^3 
                    \right] -
        \frac{1}{9} \left [ \left( \frac{\sigma}{x} \right)^9  - 
		                \left( \frac{\sigma}{y} \right)^9 
                    \right] -
\nonumber \\ & &
\nonumber \\ & &
        \frac{1}{3} \left [ \left( \frac{\sigma}{r_c} \right)^{12}  - 
	     	                \left( \frac{\sigma}{r_c} \right)^{6}  
                    \right]
			  \left [ \left( \frac{x}{\sigma} \right)^3  -     
			          \left( \frac{y}{\sigma} \right)^3
                    \right]
               \Bigg \} \frac{\sigma^3}{d^3}
\end{eqnarray} 

The compressibility factor may be determined by density differentiation
of the free energy, which leads to:
\begin{equation}
Z = \frac{pV}{NkT} = Z_0 + Z_1 + Z_2
\end{equation}
where
\begin{equation}
Z_0 = \frac{1+\eta + \eta^2 - \eta^3}{(1-\eta)^3}
\end{equation}
\begin{eqnarray} 
Z_1 & = & a_1 - \frac{12\eta^2\beta\epsilon}{d^3}
           \Bigg [ k_1
	     \left ( \frac{(5/2+\eta/2)z_1d+4+2\eta}
	                  {z_1^2(1-\eta)^3Q(z_1d)}    -
                   \frac{L(z_1d)Q_{\eta}'(z_1d)}
			      {z_1^2(1-\eta)^2Q^2(z_1d)}
           \right)  
\nonumber \\ & &
\nonumber \\ & & \qquad \qquad
	          - k_2
	     \left ( \frac{(5/2+\eta/2)z_2d+4+2\eta}
	                  {z_2^2(1-\eta)^3Q(z_2d)}    -
                   \frac{L(z_2d)Q_{\eta}'(z_2d)}
			      {z_2^2(1-\eta)^2Q^2(z_2d)}
           \right) 
	     \Bigg ] 
\nonumber \\ & &
\nonumber \\ & &  \qquad \qquad \qquad -
	     16\eta\beta\epsilon d
	         \frac{\partial g_0(d)}{\partial d}
                              W_{\rm cs}(t\sigma,d) 
\end{eqnarray}
\begin{eqnarray}
Z_2 & = & a_2 + \frac{12\eta^2\beta^2\epsilon^2}{d^3}
          \Bigg [
	          \frac{k_1^2Q_{\eta}'(z_1d)}{z_1Q^5(z_1d)} +
	          \frac{k_2^2Q_{\eta}'(z_2d)}{z_2Q^5(z_2d)} 
\nonumber \\ & &		    
\nonumber \\ & &  \qquad \qquad  \qquad -    
    \frac{2k_1k_2[Q_{\eta}'(z_1d)Q(z_2d)+Q(z_1d)Q_{\eta}'(z_2d)}
         {        (z_1+z_2)Q^3(z_1d)Q^3(z_2d)                  }
	    \Bigg ] 
\nonumber \\ & &		    
\nonumber \\ & &  +    
	    48\eta^2\beta^2\epsilon^2 
	    \left [ 
	            \frac{k_1/d}{Q^3(z_1d)}Q_{\eta}'(z_1d) -
	            \frac{k_2/d}{Q^3(z_2d)}Q_{\eta}'(z_2d) 
          \right ] W_{\rm cs}(t\sigma,d)
\end{eqnarray}
while
\begin{equation} 
d \frac{\partial g_0(d)}{\partial d} =
   \frac{3\eta(5+\eta)}{2(1-\eta)^3}
\end{equation}
and
\begin{equation} 
Q_{\eta}'(t) = \frac{6(1-\eta)t^2+36\eta t-12(1+5\eta)+12
                               [(1+2\eta)t+1+5\eta]e^{-t} }
                    {(1-\eta)^3t^3}
\end{equation} 

In order to calculate the associating strength, $\delta$, the
radial distribution function could have been assumed to be
$g=g_0+g_1$, which is already a rather good approximation. \cite{tang97a}
However, we use the SEXP (simplified Exponential) approximation
which considerably improves the estimate of $g$ around $\sigma$ with
no additional information. \cite{tang97b} According to this approximation,
$g=g_0 e^{g_1}$. Once $g$ is known, $\delta$ is calculated directly
by invoking Eq.\ \ref{eqdelta}. The integral must be performed
numerically but would have been analytic if the bond length was
held fixed.


\begin{references}


\bibitem{prigogine57}
I. Prigogine, {\em The Molecular Theory of Solutions} (North-Holland,
  Amsterdam, 1957).

\bibitem{flory64a}
P.~J. Flory, R.~A. Orwoll, and A. Vrij, J.Am.Chem.Soc. {\bf 86,} 3507 (1964).

\bibitem{flory64b}
P.~J. Flory, R.~A. Orwoll, and A. Vrij, J.Am.Chem.Soc. {\bf 86,} 3515 (1964).

\bibitem{flory64c}
P.~J. Flory, J.Am.Chem.Soc. {\bf 87,} 1833 (1964).

\bibitem{flory41}
P.~J. Flory, J.Chem.Phys. {\bf 9,} 660 (1941).

\bibitem{huggins41}
M.~L. Huggins, J.Chem.Phys. {\bf 9,} 440 (1941).

\bibitem{sanchez76}
I.~C. Sanchez and R.~H. Lacombe, J.Phys.Chem. {\bf 80,} 2352 (1976).

\bibitem{curro87}
J.~G. Curro and K.~S. Schweizer, J.Chem.Phys. {\bf 87,} 1842 (1987).

\bibitem{honnell89}
K.~G. Honnell and C.~K. Hall, J.Chem.Phys. {\bf 90,} 1841 (1989).

\bibitem{wertheim87}
M.~S. Wertheim, J.Chem.Phys. {\bf 87,} 7323 (1987).

\bibitem{wertheim84a}
M.~S. Wertheim, J.Stat.Phys. {\bf 35,} 19 (1984).

\bibitem{wertheim84b}
M.~S. Wertheim, J.Stat.Phys. {\bf 35,} 35 (1984).

\bibitem{wertheim86a}
M.~S. Wertheim, J.Stat.Phys. {\bf 42,} 459 (1986).

\bibitem{wertheim86b}
M.~S. Wertheim, J.Stat.Phys. {\bf 42,} 477 (1986).

\bibitem{chapman88}
W.~G. Chapman, G. Jackson, and K.~E. Gubbins, Molec.Phys. {\bf 65,} 1057
  (1988).

\bibitem{jackson95}
G. Jackson and R. Sears,  in {\em Observation, Prediction and Simulation of
  Phase Transitions in Complex Fluids}, M. Baus, L.~F. Rull, and J.-P.
  Ryckaert, eds., (Kluwer, Dordrecht, 1994), \ pp.\ 625--640.

\bibitem{chapman90}
W.~G. Chapman, J.Chem.Phys. {\bf 93,} 4299 (1990).

\bibitem{chapman90b}
W.~G. Chapman, K. Gubbins, G. Jackson, and M. Radosz, Ind.Eng.Chem.Res. {\bf
  29,} 1709 (1990).

\bibitem{banaszak94}
M. Banaszak, R. O'Lenick, Y.~C. Chiew, and M. Radosz, J.Chem.Phys. {\bf 100,}
  3803 (1994).

\bibitem{johnson94}
J.~K. Johnson, E.~A. Muller, and K.~E. Gubbins, J.Phys.Chem. {\bf 98,} 6413
  (1994).

\bibitem{escobedo96}
F.~A. Escobedo and J.~J. de~Pablo, Molec.Phys. {\bf 87,} 347 (1996).

\bibitem{blas97}
F.~J. Blas and L.~F. Vega, Molec.Phys. {\bf 92,} 135 (1997).

\bibitem{banaszak93}
M. Banaszak, Y.~C. Chiew, and M. Radosz, Phys.Rev.E {\bf 48,} 3760 (1993).

\bibitem{gil-villegas97}
A. Gil-Villegas, A. Galindo, P.~J. Whitehead, S.~J. Mills, G. Jackson, and
  A.~N. Burgess, J.Chem.Phys. {\bf 106,} 4168 (1997).

\bibitem{huang90}
S.~H. Huang and M. Radosz, Ind.Eng.Chem.Res. {\bf 29,} 2284 (1990).

\bibitem{chen98}
C.-K. Chen, M. Banaszak, and M. Radosz, J.Phys.Chem. {\bf B-102,} 2427 (1998).

\bibitem{blas98}
F.~J. Blas and L.~F. Vega, Ind.Eng.Chem.Res. {\bf 37,} 660 (1998).

\bibitem{vega94}
C. Vega, S. Lago, and B. Garzon, J.Chem.Phys. {\bf 100,} 2182 (1994).

\bibitem{vega96b}
C. Vega, L.~G. MacDowell, and P. Padilla, J.Chem.Phys. {\bf 104,} 701 (1996).

\bibitem{chatterjee98}
A.~P. Chatterjee and K.~S. Schweizer, J.Chem.Phys. {\bf 108,} 3813 (1998).

\bibitem{rowlinson82}
J.~S. Rowlinson and F.~L. Swinton, {\em Liquids and Liquid Mixtures}, 3rd  ed.
  (Butterworth, London, 1982).

\bibitem{tsonopoulos87}
C. Tsonopoulos, AIChE. J. {\bf 33,} 2080 (1987).

\bibitem{tsonopoulos93}
C. Tsonopoulos and Z. Tan, Fluid Phase Equilibria {\bf 83,} 127 (1993).

\bibitem{nakanishi60}
K. Nakanishi, M. Kurata, and M. Tamura, J.Chem.Eng.Data {\bf 5,} 210 (1960).

\bibitem{korsten98}
H. Korsten, Chem.Eng.Technol. {\bf 21,} 229 (1998).

\bibitem{dup82}
B. Duplantier, J.Phys. (France) {\bf 43,} 911 (1982).

\bibitem{dup87}
B. Duplantier, J.Chem.Phys. {\bf 86,} 4233 (1987).

\bibitem{hs99}
J. Hager and L. Sch{\"a}fer, Phys.Rev.E {\bf 60,} 2071 (1999).

\bibitem{anselme90}
M.~J. Anselme, M. Gude, and A.~S. Teja, Fluid Phase Equilibria {\bf 57,} 317
  (1990).

\bibitem{nikitin94}
E.~D. Nikitin, P.~A. Pavlov, and N.~V. Bessonova, J.Chem.Thermodyn. {\bf 26,}
  177 (1994).

\bibitem{nikitin97}
E.~D. Nikitin, P.~A. Pavlov, and A.~P. Popov, Fluid Phase Equilibria {\bf 141,}
  155 (1997).

\bibitem{smit95a}
B. Smit, S. Karaborni, and J.~I. Siepmann, J.Chem.Phys. {\bf 102,} 2126 (1995),
  erratum: J.Chem.Phys. {\bf 109}, 352 (1998).

\bibitem{sheng94}
Y.-J. Sheng, A.~Z. Panagiotopoulos, S.~K. Kumar, and I. Szleifer,
  Macromolecules {\bf 27,} 400 (1994).

\bibitem{wilding96}
N.~B. Wilding, M. Muller, and K. Binder, J.Chem.Phys. {\bf 105,} 802 (1996).

\bibitem{vega96a}
C. Vega and L.~G. MacDowell, Molec.Phys. {\bf 88,} 1575 (1996).

\bibitem{nikitin96}
E.~D. Nikitin, P.~A. Pavlov, and P.~V. Skripov, Int.J.Thermophys. {\bf 17,} 455
  (1996).

\bibitem{zhou92}
Y. Zhou and G. Stell, J.Chem.Phys. {\bf 96,} 1507 (1992).

\bibitem{friedman85}
H.~L. Friedman, {\em A Course in Statistical Mechanics} (Prentice-Hall,
  Englewood Cliffs, 1985).

\bibitem{smith98}
W.~R. Smith, I. Nezbeda, M. Strnad, B. Triska, S. Labik, and A. Malijevsky,
  J.Chem.Phys. {\bf 109,} 1052 (1998).

\bibitem{ghonasgi94}
D. Ghonasgi and W.~G. Chapman, AIChE. J. {\bf 40,} 878 (1994).

\bibitem{lado73}
F. Lado, Phys.Rev. {\bf A-8,} 2548 (1973).

\bibitem{lado83}
F. Lado, S.~M. Foiles, and N.~W. Ashcroft, Phys.Rev. {\bf A-45,} 2374 (1983).

\bibitem{labik85}
S. Labik, A. Malijevsky, and P. Vonka, Molec.Phys. {\bf 56,} 709 (1985).

\bibitem{mcquarrie76}
D.~A. McQuarrie, {\em Statistical Mechanics} (Harper \& Row, New York, 1976).

\bibitem{barker67}
J.~A. Barker and D. Henderson, J.Chem.Phys. {\bf 47,} 4714 (1967).

\bibitem{weeks71}
J.~D. Weeks, D. Chandler, and H.~C. Andersen, J.Chem.Phys. {\bf 54,} 5237
  (1971).

\bibitem{tang97c}
Y. Tang and B.~C.-Y. Lu, Molec.Phys. {\bf 90,} 215 (1997).

\bibitem{tang97a}
Y. Tang, Z. Tong, and B.~C.-Y. Lu, Fluid Phase Equilibria {\bf 134,} 21 (1997).

\bibitem{tang97b}
Y. Tang and B.~C.-Y. Lu, AIChE. J. {\bf 43,} 2215 (1997).

\bibitem{tang93}
Y. Tang and B.~C.-Y. Lu, J.Chem.Phys. {\bf 99,} 9828 (1993).

\bibitem{smit95b}
B. Smit, Molec.Phys. {\bf 85,} 153 (1995).

\bibitem{frenkel96}
D. Frenkel and B. Smit, {\em Understanding Molecular Simulation} (Academic
  Press, San Diego, 1957).

\bibitem{mw94}
M. M{\"u}ller and N.~B. Wilding, Phys.Rev.E {\bf 57,} 2076 (1994).

\bibitem{mueller99}
M. M{\"u}ller and L.~G. MacDowell, Submitted to Macromolecules  .

\bibitem{johnson93}
J.~K. Johnson, J.~A. Zollweg, and K.~E. Gubbins, Molec.Phys. {\bf 78,} 591
  (1993).

\bibitem{malijevsky87}
A. Malijevsky and S. Labik, Molec.Phys. {\bf 60,} 663 (1987).

\bibitem{labik89}
S. Labik and A. Malijevsky, Molec.Phys. {\bf 67,} 431 (1989).

\bibitem{hansen86}
J.-P. Hansen and I.~R. McDonald, {\em Theory of Simple Liquids} (Academic
  Press, London, 1976).

\bibitem{desouza93}
L.~E.~S. de~Souza and D. Ben-Amotz, Molec.Phys. {\bf 78,} 137 (1993).


\end{references}

\newpage

\begin{figure}
         \caption{
         \label{fig:cycle} 
         Relating the structure of the fluid to the excess chemical
         potential of the components by means of a thermodynamic
         cycle.
         }
\end{figure}

\begin{figure}
         \caption{
            \label{fig:mapping} 
	    Mapping of the probability distribution of the density onto 
	    the universal
	    distribution of the 3D Ising universality class (line).
	    The chain length $n$ and the estimate for the critical temperature
	    are indicated in the key.
         }
\end{figure}

\begin{figure}
         \caption{
            \label{fig:pressure} 
             Pressure against monomer density for chains of 10 monomers.
             All quantities are given in Lennard-Jones reduced units. 
		 Symbols are NVT simulation data while lines are predictions
             from TPT1; full line, RHNC version; dashed line, MSA
		 version. From top to bottom, pressure isotherms at $T=5,
		 4, 3, 2.5$ and $1.68$ reduced LJ units.
         }
\end{figure}

\begin{figure}
         \caption{
            \label{fig:chempot} 
             Excess chemical potential against monomer density for
             chains of 10 monomers. All quantities are given in Lennard--Jones 
	     reduced units. Symbols are Grand Canonical simulation
	     data, while lines are predictions from TPT1;
	     full line, RHNC version; dashed line, MSA version.
	     From top to bottom, chemical potential isotherms at $T=5,
	     4, 3, 2.5$ and $1.68$ LJ reduced units.
         }
\end{figure}

\begin{figure}
         \caption{
            \label{fig:phasen} 
                 Liquid--vapor coexistence curves of a 10-mer as obtained from 
		 grandcanonical simulations (solid lines) and NpT simulations 
		 at $p=0$ (diamonds), compared with TPT1-RHNC (dashed line) and 
		 TPT1-MSA (dotted line). The filled circle presents the critical 
		 point as extracted from finite size scaling of the MC data.
		 The open circle and the open square denote the critical point
		 of the TPT1-RHNC and the TPT1-MSA, respectively. The simulation 
             results
		 and TPT1-MSA calculations for monomers ($n=1$) are also included.
         }
\end{figure}

\begin{figure}
         \caption{
            \label{fig:critical point} 
	    Critical temperatures ({\bf a}) and critical monomer densities 
	     ({\bf b})
	    in the MC simulations and the perturbation theory. In panel ({\bf a})
	    the temperatures of the intersections of $R_e^2(T)/(n-1)$
	    for neighboring chain lengths are also included. For $n \to \infty$
	    the values tend to the $\Theta$ temperature.
         }
\end{figure}

\begin{figure}
         \caption{
            \label{fig:theta} 
	    Temperature dependence of end--to--end distance of a single chain
	    in the vicinity of the $\Theta$ temperature. The crossings of the
	    ratios $R_e^2(T)/(n-1)$ for neighboring chain lengths are indicated 
	    by arrows. The crossing points converge to the $\Theta$ temperature 
	    in the limit of infinite chain length $n \to \infty$.
         }
\end{figure}

\newpage

\begin{table} 
   \begin{tabular}{|c|ccc|}
   \hline
     $n$  &  $T_c$ (MC)  &  $T_c$ (RHNC)    &   $T_c$ (MSA)    \\
    \hline
      1  & 1.00   &   1.02       &      1.11   \\
     10  & 1.98   &    2.27  &      2.36   \\            
     20  & 2.214  &    2.56  &      2.62   \\
     40  & 2.396  &    2.79  &      2.81   \\
     60  & 2.485  &    2.90  &      2.88    \\
$\infty$ & $\sim$ 3.3  &  3.44   &  3.14  \\
   \hline
     $n$  & $n \rho_c$ (MC)  &    $n\rho_c$ (RHNC)  &  $n \rho_c$ (MSA)    \\
      1 & 0.321  &    0.376  &    0.323    \\ 
     10 & 0.245  &    0.207   &    0.217    \\
     20 & 0.206  &    0.145   &    0.184    \\
     40 & 0.172  &    0.108   &    0.150     \\
     60 & 0.1523 &    0.091   &    0.140    
  \end{tabular}
\caption{Critical temperature, $T_c$ and critical monomer density,
	   $n\rho_c$ as obtained from simulation (MC) and from TPT1
	   with either the RHNC version or the MSA version for the
	   structure and thermodynamics of the reference fluid}
\label{tab:cripoint}
\end{table}

\begin{table} 
\begin{displaymath}
\begin{array}{lll}
\hline
i &  \qquad \qquad  a_i     &  \qquad \qquad       b_i          \\
\hline
1 & x_1T + x_2 \surd T + x_3 + x_4 /T + x_5/T^2  
  & x_{20}/T^2+x_{21}/T^3                            \\
2 & x_6T + x_7 + x_8/T + x_9/T^2
  & x_{22}/T^2+x_{23}/T^4                            \\
3 & x_{10}T + x_{11} + x_{12}/T
  & x_{24}/T^2+x_{25}/T^3                            \\
4 & x_{13}
  & x_{26}/T^2+x_{27}/T^4                            \\
5 & x_{14}/T + x_{15}/T^2
  & x_{28}/T^2+x_{29}/T^3                            \\
6 & x_{16}/T
  & x_{30}/T^2+x_{31}/T^3+x_{32}/T^4                 \\
7 & x_{17}/T + x_{18}/T^2       &   {\rm -}          \\
8 & x_{19}/T^2                  &   {\rm -}          \\
\hline
\end{array}
\end{displaymath}
\caption{The $a_i$ and $b_i$ temperature dependent coefficients of the
          BWR equation of state (\protect{\ref{eq:bwr}}). 
	    The $x_j$ are adjustable
	    parameters whose actual value are given in Table
	    \protect{\ref{tab:parbwr}}.  }
\label{tab:coefbwr}
\end{table}

\begin{table}  
   \begin{tabular}{|cccc|}
   \hline
   i       &     $x_i$             & i  &   $x_i$        \\
   \hline
    1       &  .421192000D+00   & 17 & -.107917493D+03    \\
    2       &  .476645300D+01   & 18 &  .946452266D+03   \\
    3       & -.915420500D+01   & 19 & -.661602970D+03   \\
    4       &  .161011000D+01   & 20 &  .117245702D+03  \\
    5       & -.152328700D+01   & 21 & -.348234700D+01   \\
    6       &  .205245400D+01   & 22 &  .415445097D+03  \\
    7       & -.348800300D+01   & 23 &  .325092260D+02   \\
    8       &  .459058600D+01   & 24 &  .136021606D+04  \\
    9       & -.111109576D+03   & 25 & -.672466920D+03   \\
   10       & -.683045000D+00   & 26 & -.114228784D+03   \\
   11       &  .106678670D+02   & 27 &  .602317138D+03   \\
   12       & -.243251130D+02   & 28 &  .185990661D+03   \\
   13       &  .193579400D+02   & 29 & -.116953259D+04  \\
   14       & -.196790929D+03   & 30 &  .607836000D+00   \\
   15       & -.172144369D+03   & 31 &  .656689000D+00  \\
   16       &  .262199061D+03   & 32 &  .588415000D+00 \\
   \hline
   \end{tabular}
\caption{Parameters for the fit of the RHNC data to the 
	   Modified Benedict-Webb-Rubin equation of state. Notation
	   as in paper \protect\cite{johnson93}. The nonlinear parameter is set
	   to \protect{$\gamma=3$}
	   }
\label{tab:parbwr}
\end{table}

\begin{table} 
    \begin{tabular}{|cccccc|}
 \hline
 i  &       j=1          &       j=2          &       j=3          &
            j=4          &       j=5          \\    
 \hline
 1  &  .496554000D+00 &  .129871000D+01 & -.595011800D+01 &
       .860280100D+01 & -.452759900D+01 \\
 2  &  .101561200D+01 & -.967002900D+01 &  .677884460D+02 &
      -.950135670D+02 &  .471191210D+02 \\
 3  & -.604253400D+01 &  .411972640D+02 & -.252396610D+03 &
       .376561440D+03 & -.178249974D+03 \\
 4  &  .622698400D+01 & -.680522730D+02 &  .391425109D+03 &
      -.605492785D+03 &  .292582291D+03 \\
 5  & -.691995000D+00 &  .357118100D+02 & -.218469635D+03 &
       .348511635D+03 & -.171957868D+03 \\ 
 \hline
    \end{tabular}
\caption{Parameters $a_{ij}$ for the fit to \protect{$\delta/\delta_0$}
         (Eq. \protect\ref{eqfit_delta}) as obtained from the RHNC 
         integral equation}
\label{tab:pargr}
\end{table}

\newpage

\begin{figure}[htbp]
    \begin{minipage}[t]{160mm}%
           \mbox{
	           \setlength{\epsfxsize}{8cm}
		           \epsffile{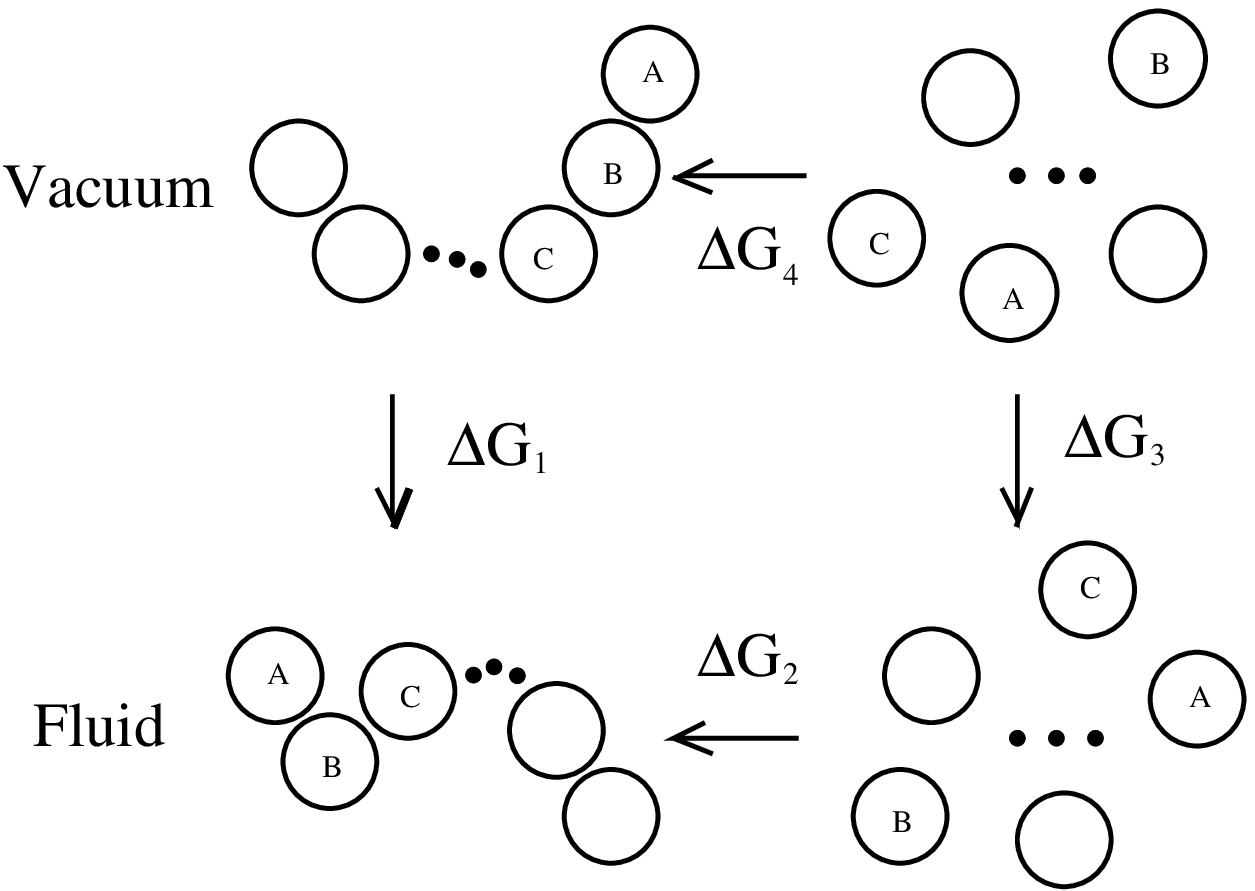}
			          }
    \end{minipage}%
\end{figure}
\vspace{5cm}
figure 1

\newpage

\begin{figure}[htbp]
    \begin{minipage}[t]{160mm}%
           \mbox{
	           \setlength{\epsfxsize}{8cm}
		           \epsffile{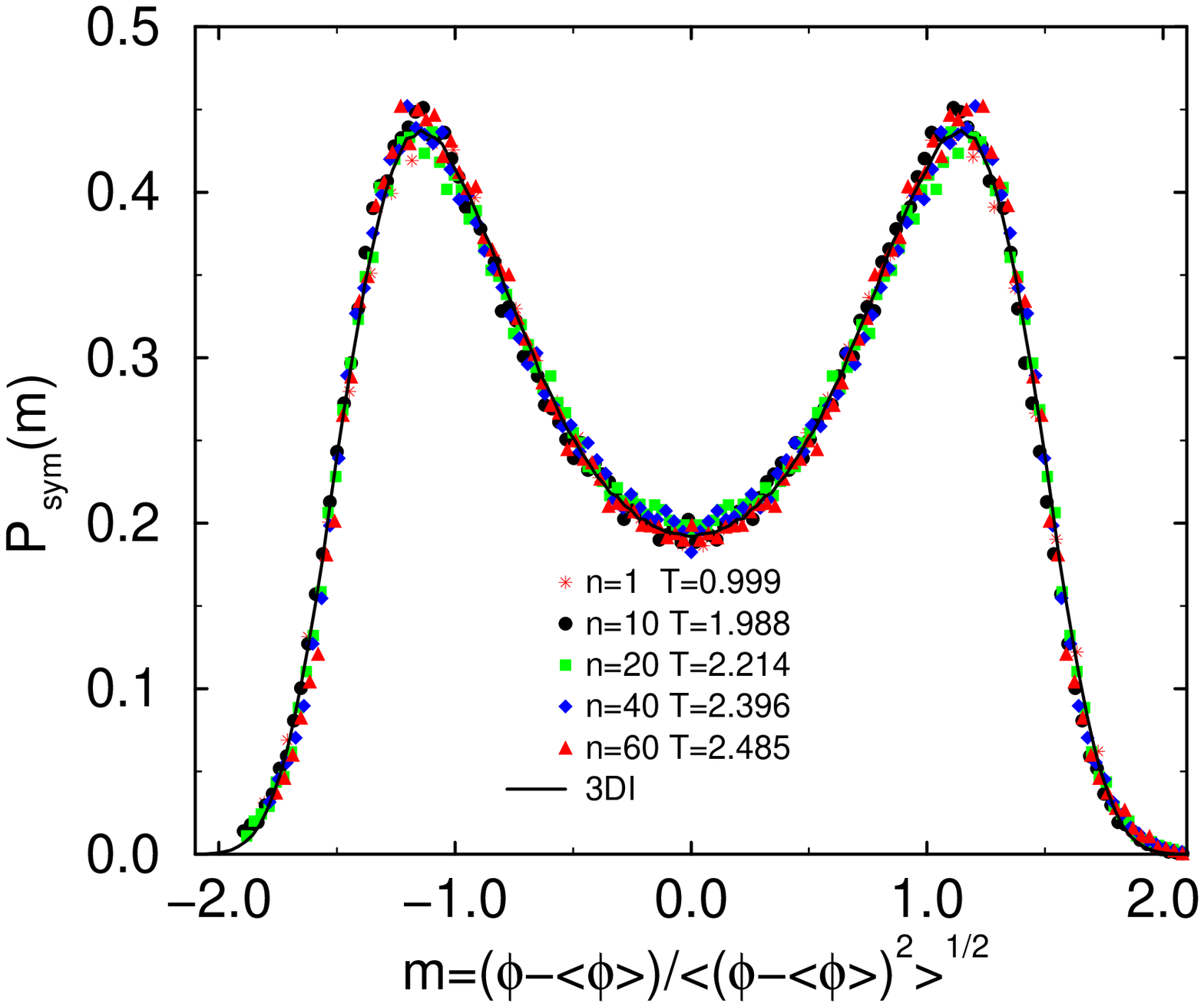}
			          }
    \end{minipage}%
\end{figure}
\vspace{5cm}
figure 2

\newpage

\begin{figure}[htbp]
    \begin{minipage}[t]{160mm}%
           \mbox{
	           \setlength{\epsfxsize}{8cm}
		           \epsffile{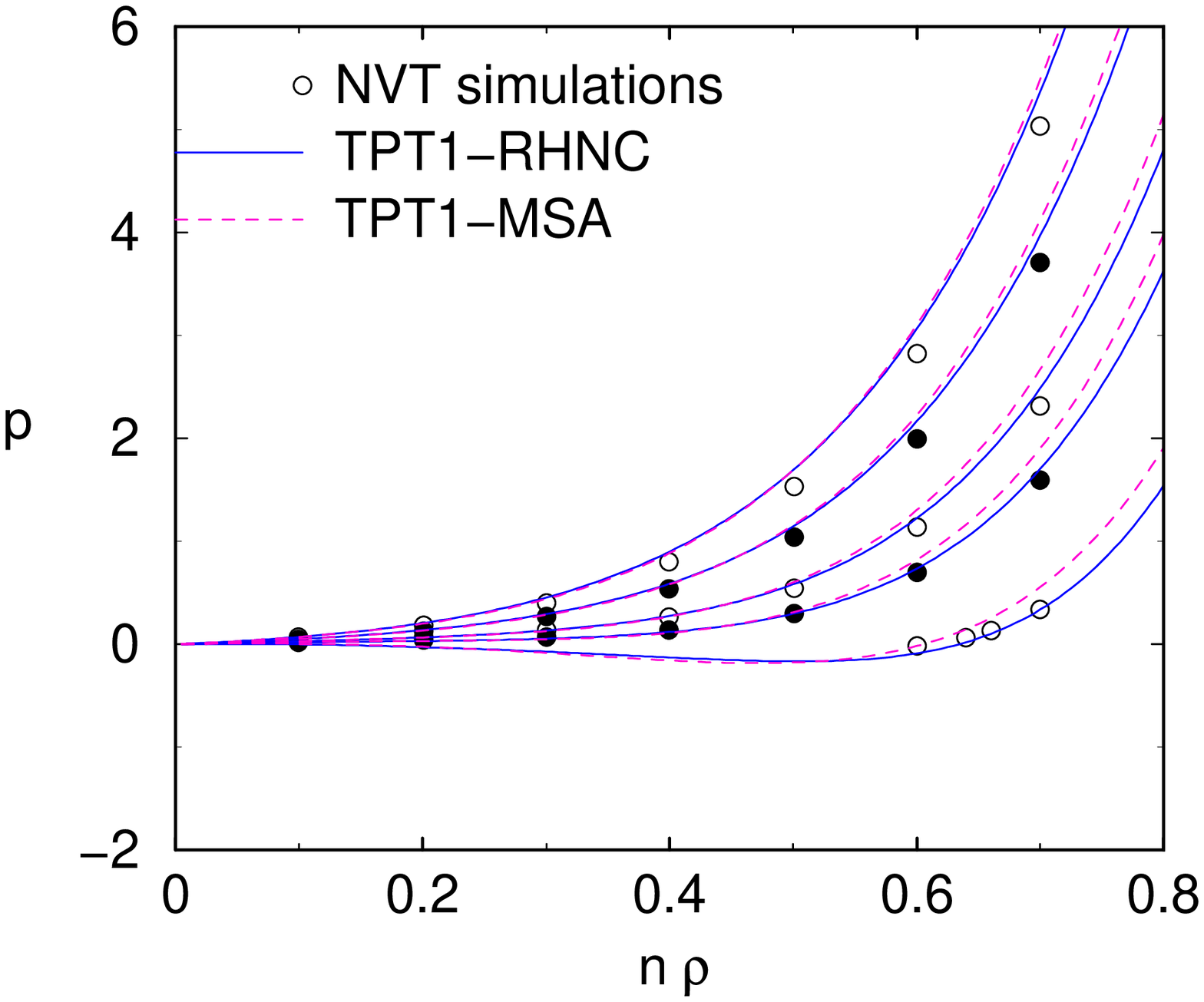}
			          }
    \end{minipage}%
    \hfill%
\end{figure}
\vspace{5cm}
figure 3

\newpage

\begin{figure}[htbp]
    \begin{minipage}[t]{160mm}%
           \mbox{
	           \setlength{\epsfxsize}{8cm}
		           \epsffile{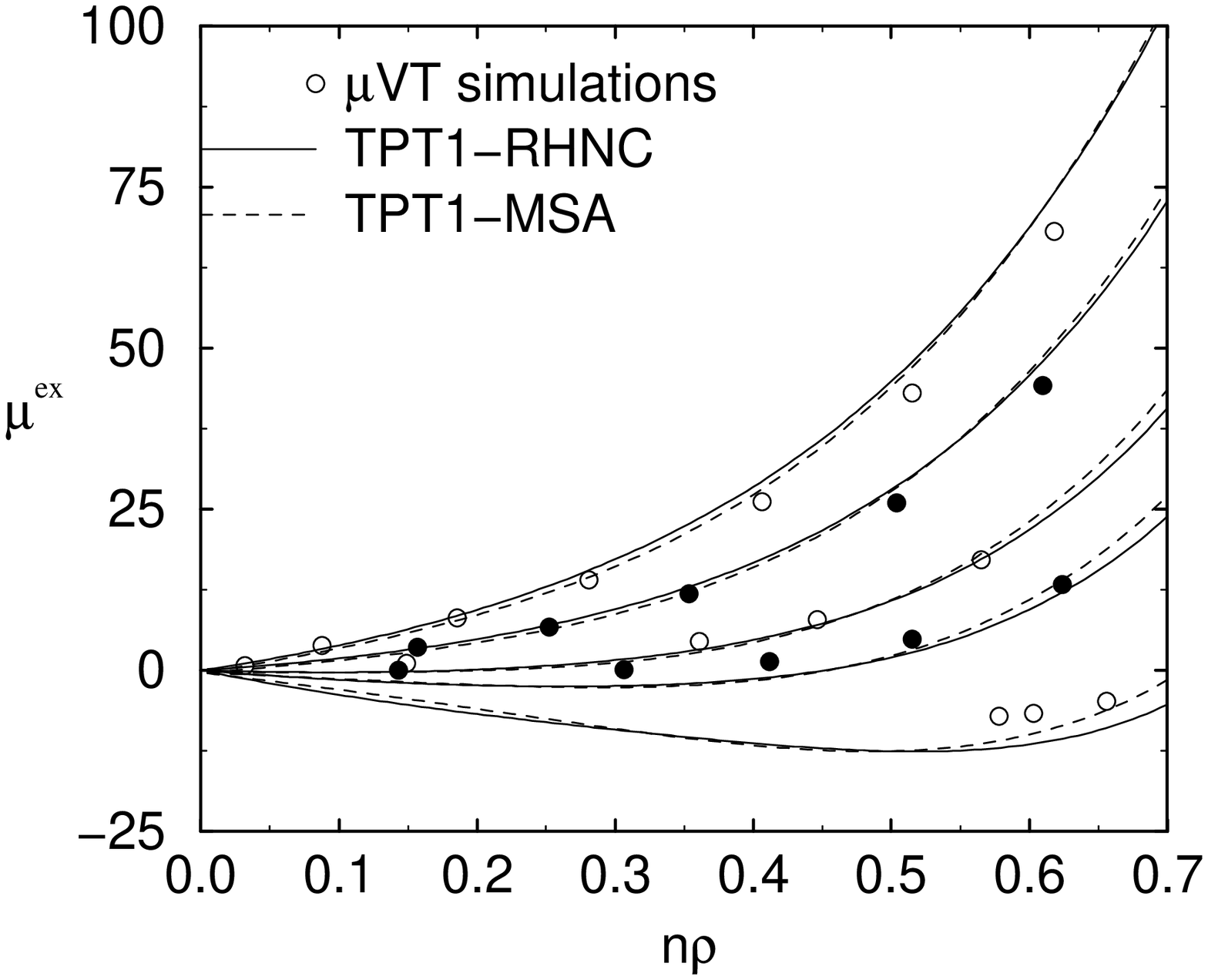}
			          }
    \end{minipage}%
    \hfill%
\end{figure}
\vspace{5cm}
figure 4

\newpage

\begin{figure}[htbp]
    \begin{minipage}[t]{160mm}%
           \mbox{
	           \setlength{\epsfxsize}{8cm}
		           \epsffile{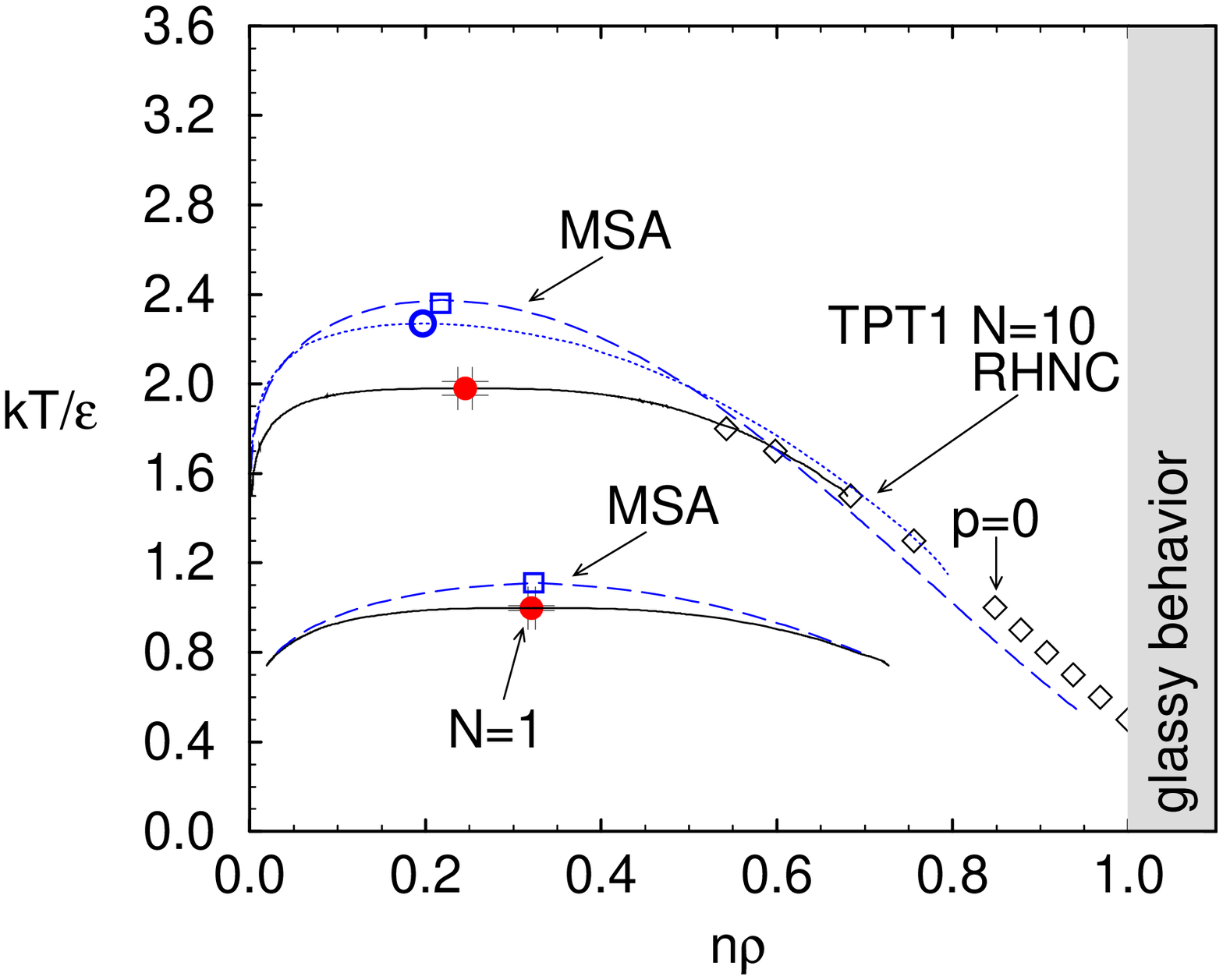}
			          }
    \end{minipage}%
\end{figure}
\vspace{5cm}
figure 5

\newpage

\begin{figure}[htbp]
    \begin{minipage}[t]{160mm}%
           \mbox{
           \hspace*{-1.0cm}
	   ({\bf a})
	           \setlength{\epsfxsize}{7cm}
		           \epsffile{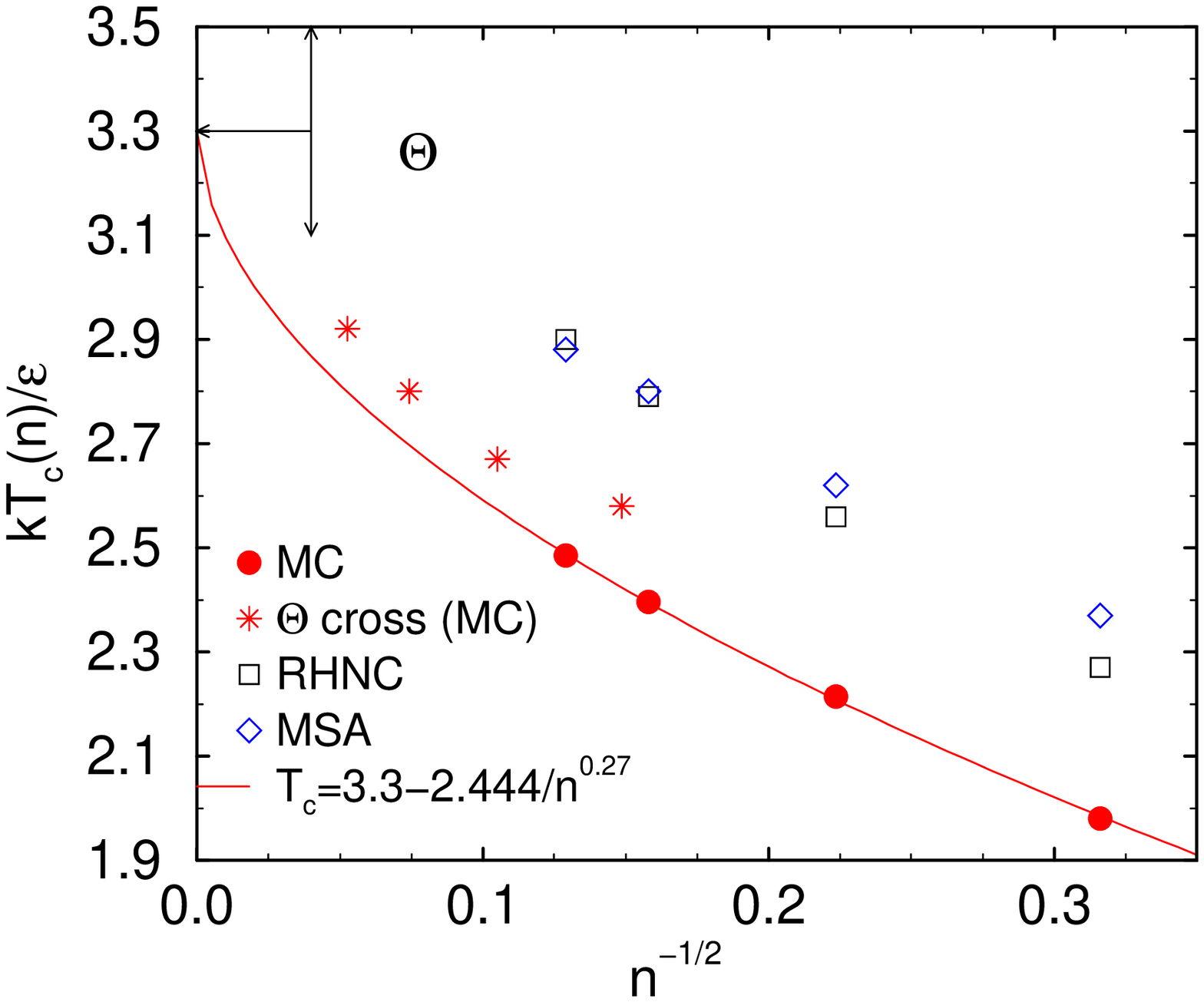}
           \hspace*{0.3cm}
	   ({\bf b})
           \setlength{\epsfxsize}{7cm}
		           \epsffile{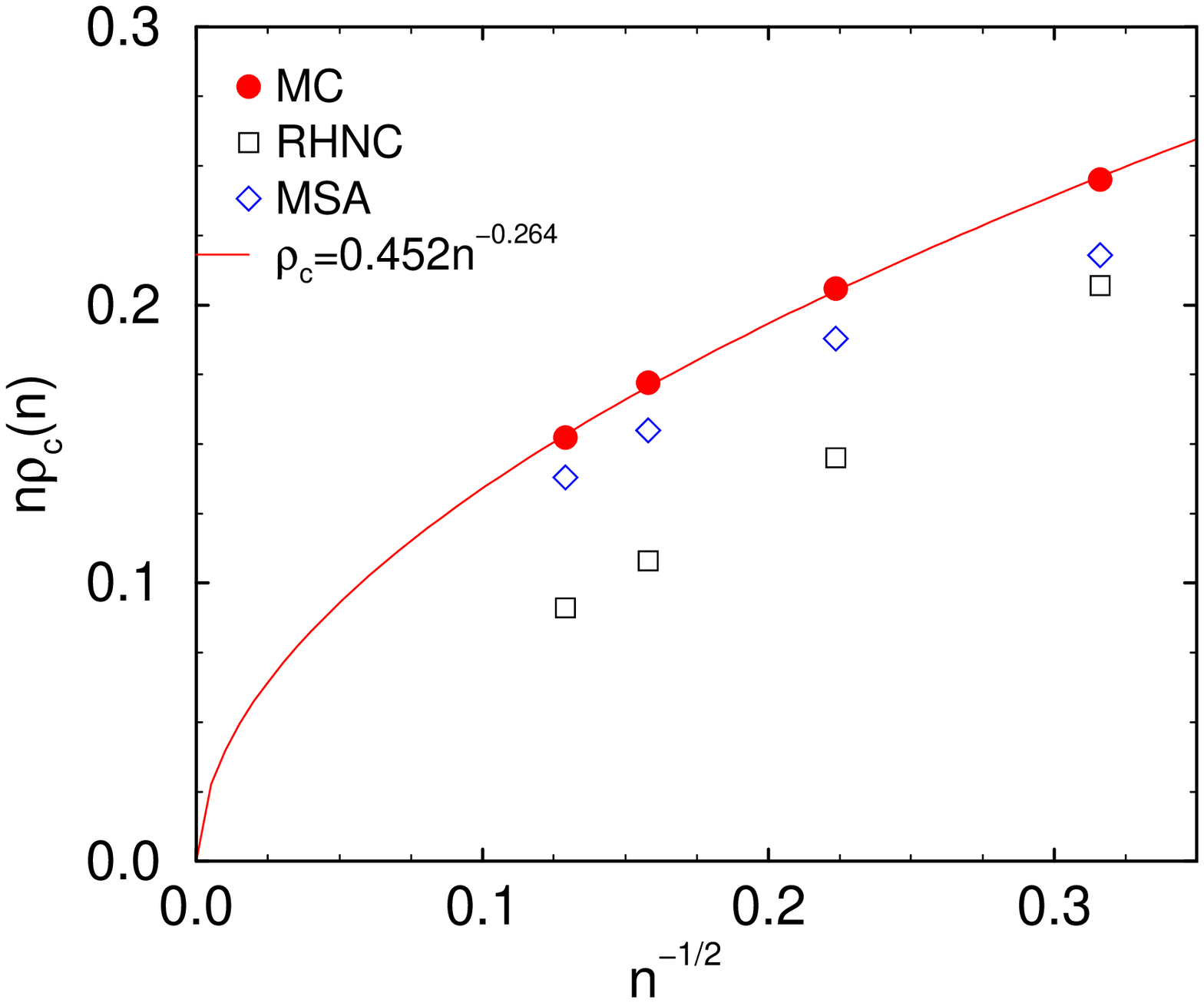}
			          }
    \end{minipage}%
\end{figure}
\vspace{5cm}
figure 6

\newpage

\begin{figure}[htbp]
    \begin{minipage}[t]{160mm}%
           \mbox{
	           \setlength{\epsfxsize}{8cm}
		           \epsffile{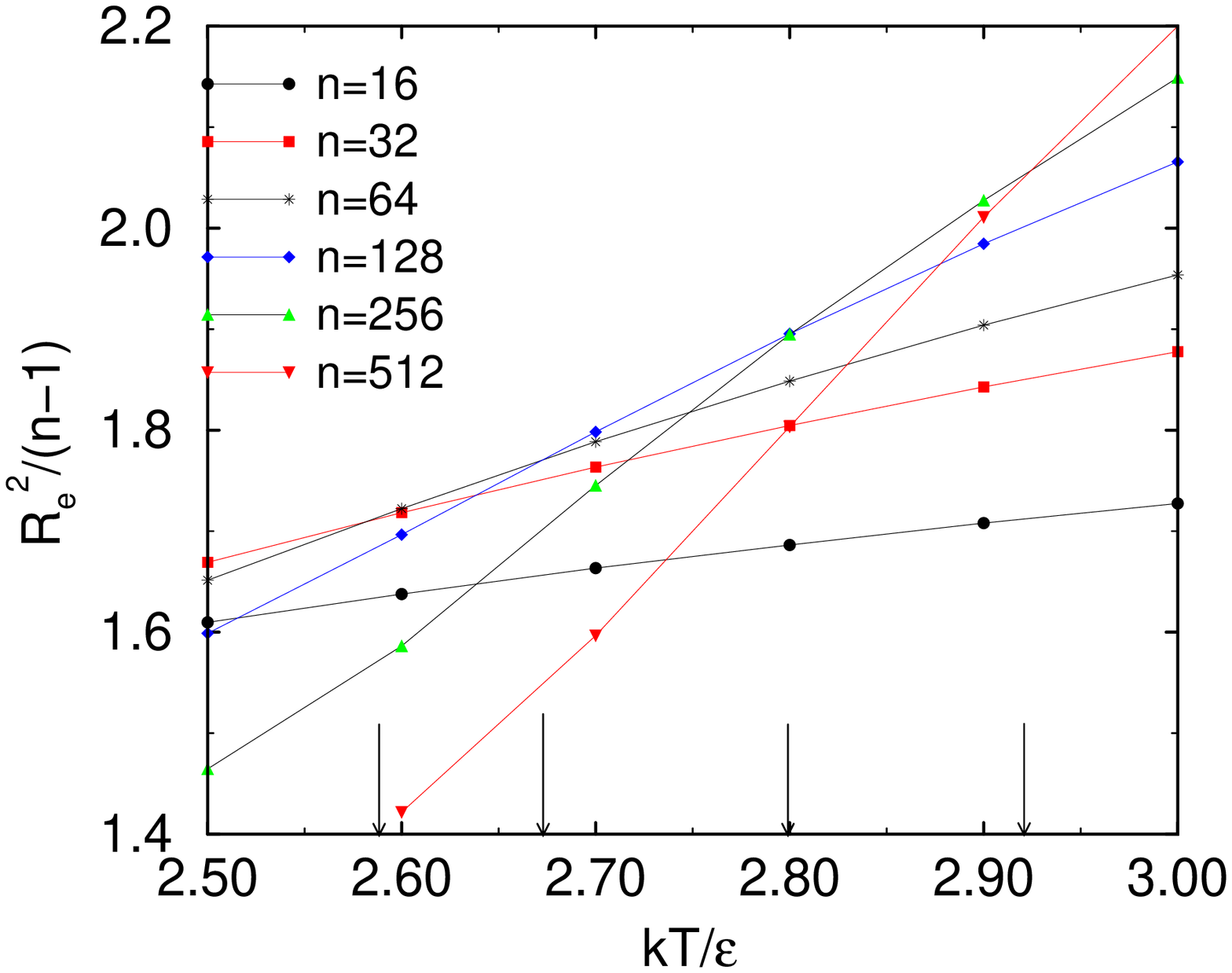}
			          }
    \end{minipage}%
\end{figure}
\vspace{5cm}
figure 7

\end{document}